# STANDARD MODEL PROCESSES

*Conveners* : F. Boudjema[1] and B. Mele[2]

*Working group* :
E. Accomando[3], S. Ambrosanio[2], A. Ballestrero[3], D. Bardin[4], G. Bélanger[1], F. Berends[5], M. Bonesini[6], E. Boos[7], M. Cacciari[8], F. Caravaglios[9], M. Dubinin[7], J. Fujimoto[10], E. Gabrielli[2], A. Hasan[11], W. Hollik[12], T. Ishikawa[10], S. Jadach[13], T. Kaneko[14], K. Kato[15], S. Kawabata[10], R. Kleiss[16], Y. Kurihara[10], D. Lehner[17], R. Miquel[18], K. Moenig[18], G. Montagna[19], M. Moretti[20], O. Nicrosini[18], G.J. van Oldenborgh[5], C. Papadopoulos[21], J. Papavassiliou[22], G. Passarino[3], D. Perret-Gallix[1], F. Piccinini[19], R. Pittau[23], E. Poli[19], L. Pollino[19], P. Razis[24], M. Schmitt[25], D.J. Schotanus[26], Y. Shimizu[10], H. Tanaka[27], L. Trentadue[28], J. Ulbricht[29], C. Verzegnassi[30], B.F.L. Ward[31], Z. Wąs[13], G.W. Wilson[32].

**Abstract**

We present the results obtained by the Standard Model Process group in the CERN Workshop "Physics at LEP2" (1994/95).

1) ENS-LAPP, Annecy, France
2) INFN and Univ. Rome 1, Italy
3) INFN and Univ. Torino, Italy
4) DESY-IfH Zeuthen, Germany and JINR Dubna, Russia
5) Univ. Leiden, The Netherlands
6) INFN and Univ. Milan, Italy
7) Moscow State Univ., Russia
8) DESY, Hamburg, Germany
9) Univ. Oxford, United Kindom
10) KEK, Tsukuba, Japan
11) ETH, Zürich, Switzerland
12) Univ. Karlsruhe, Germany
13) CERN, Geneva, Switzerland and INP, Krakow, Poland
14) Univ. Meiji-Gakuin, Totsuka, Japan
15) Univ. Kogakuin, Shinjuku, Japan
16) NIKHEF, Amsterdam, The Netherlands
17) DESY, Zeuthen, Germany
18) CERN, Geneva, Switzerland
19) INFN and Univ. Pavia, Italy
20) Univ. Southampton, United Kindom
21) Univ. Durham, United Kindom and CERN, Geneva, Switzerland
22) CPT-Marseille, France
23) PSI, Villigen, Switzerland
24) Univ. Cyprus, Nicosia, Cyprus
25) Univ. Wisconsin, USA
26) NIKHEF and Univ. Nijmegen, The Netherlands
27) Univ. Rikkyo, Nishi-Ikebukuro, Japan
28) Univ. Parma, Italy and CERN, Geneva, Switzerland
29) ETH, Zürich, Switzerland
30) Univ. Lecce, Italy
31) Univ. Tennessee, Knoxville, TN and SLAC, Univ. Stanford, CA, USA
32) Univ. Hamburg and DESY, Hamburg, Germany.

# Contents

# 1 Introduction

While the energy increase of LEP will enable pair production of $W$'s and might open up the threshold for new particles, notably the Higgs boson, a host of standard model processes will also show up, as shown in Fig. 1.

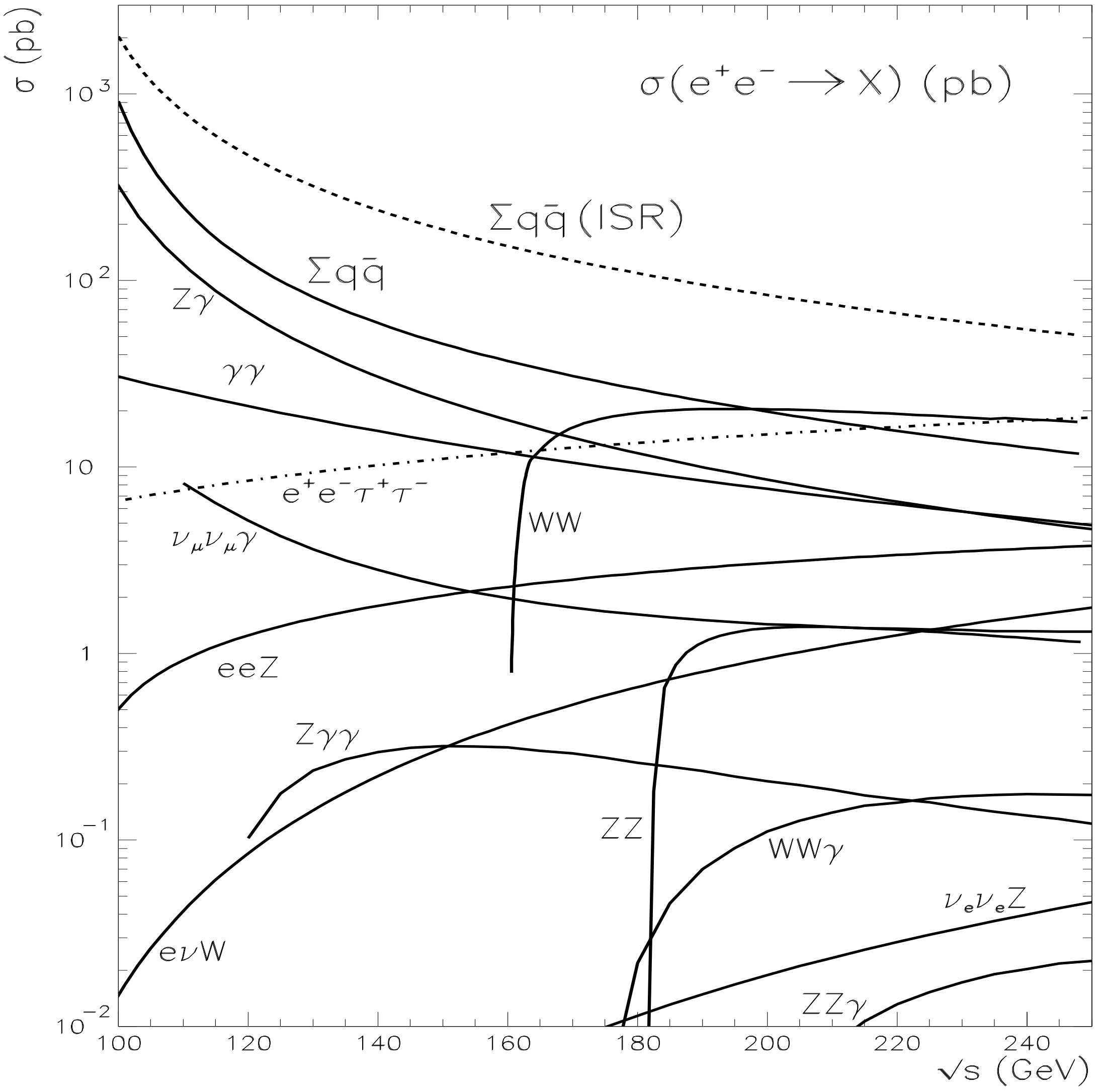

Figure 1: *Cross sections for some typical standard model processes.*
*For* $e^+e^- \rightarrow e^+e^- Z, e\nu_e W, \nu_e\bar{\nu}_e Z$ *only the dominant t-channel contribution is shown. The photons in* $Z\gamma$ *and* $\gamma\gamma$ *are such that* $|\cos_{e\gamma}| < 0.9$. *For* $\nu_\mu\bar{\nu}_\mu\gamma$ *there is the additional cut* $E_\gamma$ *>10GeV. In* $Z\gamma\gamma$, $W^+W^-\gamma$ *and* $ZZ\gamma$ *the photon cut is* $p_T^\gamma > 10GeV$ *and all particles are separated with opening angles:* $\widehat{eV} > 15^0, \widehat{VV'} > 10^0$; $V = W, Z, \gamma$.

Some of these processes can be considered as potential backgrounds to those most interesting signals LEP2 intends to investigate. For instance, there are four-fermion processes that cannot be associated with the doubly-resonant $WW$ production or with the Higgs-boson production. Therefore, it is essential to know as precisely as possible the expected yield for these processes. Quark- and lepton-pair production will be dominant reactions at LEP2 and can be exploited for precision tests in this new energy range. Moreover, starting below the threshold for $W$ pair production one sees that other processes will take place, like for instance single $Z$ and $W$ production. Beyond the $WW$ threshold, one can envisage $Z$-pair production or even triple vector-boson production, like $WW\gamma$, which involves the quartic vector couplings.
The aim of the Working Group has therefore been to provide as precise an evaluation as possible of all those processes that were not investigated within the *WW cross sections and distributions*, the $M_W$ or the *Higgs* groups in this Workshop, and which did not deal specifically with QCD issues. The other objective was to indicate, like in the case of two-fermion and single-photon processes, which interesting physics issues could be investigated.

# 2 Two-Fermion Production

## 2.1 General considerations, LEP2 *vs* LEP1

Quark- and lepton-pair production at LEP1 (and SLC) has provided one of the most stringent tests on the Standard Model ($\mathcal{SM}$) of electroweak interactions. It has also either constrained or ruled out some alternative models, especially through their virtual indirect effects. At LEP2 energies, this process still remains one of the dominant processes. For instance, as evidenced from Fig. 1, quark pair production has a larger cross section than the $WW$ process, the *bread and butter* of LEP2, and even larger after the inclusion of the initial state radiation (ISR). In view of this expected wealth of events, it is worth enquiring if the characteristics of the two-fermion observables will continue to be conducive to further tests of the $\mathcal{SM}$ and beyond. One important observation, however, is that, as one moves away from the $Z$-peak, not only fermion-pair production drops precipitously, but also the photon exchange becomes very important. The latter dominates the cross section for up-type quarks and even more so for $\mu^+\mu^-$, see Fig. 2. In particular, at $\sqrt{s} = 175GeV$ one has $\sigma_Z/\sigma_\gamma \simeq 0.27, 0.68, 3.52, 1.44$ for $\mu, u, d$ and hadrons respectively. Another critical fact is the very large "correction" due to initial-state radiation (QED). Above $\sim$ 100GeV this more than doubles the muon Born cross section, as displayed in Fig. 2. Therefore, it is essential that these corrections be controlled very precisely. At LEP1 the latter were also very important, leading to a $\sim 74\%$ *reduction* factor of the peak cross section, and were essentially due to soft-photon emission, while hard radiation (energetic collinear photons) was inhibited. Indeed, around the resonance $\Gamma_Z$ acts as a natural cut-off for hard radiation. On the other hand, away from the $Z$-peak, the fast decrease of the cross section favours the radiation of hard photons that boost the effective two-fermion centre-of-mass energy back to the $Z$ mass: this is the so-called *Z return*. Therefore, if for the inclusive two-fermion cross section one looks at the invariant mass of the fermion pairs, $m_{f\bar{f}} = \sqrt{s'} = \sqrt{sx}$, one sees that a large sample is clustered around $m_{f\bar{f}} \simeq M_Z$. The effect is quite dramatic for $\sigma_{q\bar{q}}$ where, at $\sqrt{s} = 180GeV$, about 70% of the events are "LEP1-type pairs", as one can see from Fig. 3.

Still, considering the canonical integrated LEP2 luminosity ($500pb^{-1}$), one expects to measure the various two-fermion observables with a good precision (even after discarding the Z-return events) especially if one combines the 4 experiments. For instance, at $\sqrt{s} = 175GeV$, the expected experimental precision on the muon cross section is about 1.3%, while for the hadronic* one expects 0.7% [2]. The corresponding error for the forward-backward asymmetry of muons is $\Delta A^{\mu}_{FB} \sim 0.01$ [2]. Note that this asymmetry is much larger than at LEP1. Another interesting observable is $R_b$, with an expected overall accuracy of about 2.5% [3]. These numbers should serve as benchmarks for the required accuracy on the theoretical calculations, $\epsilon^{th}$ . One should aim, at least, at a theoretical precision below half the values quoted above. For instance, for $\sigma_h$ one needs $\epsilon^{th}_h < 0.3\%$.

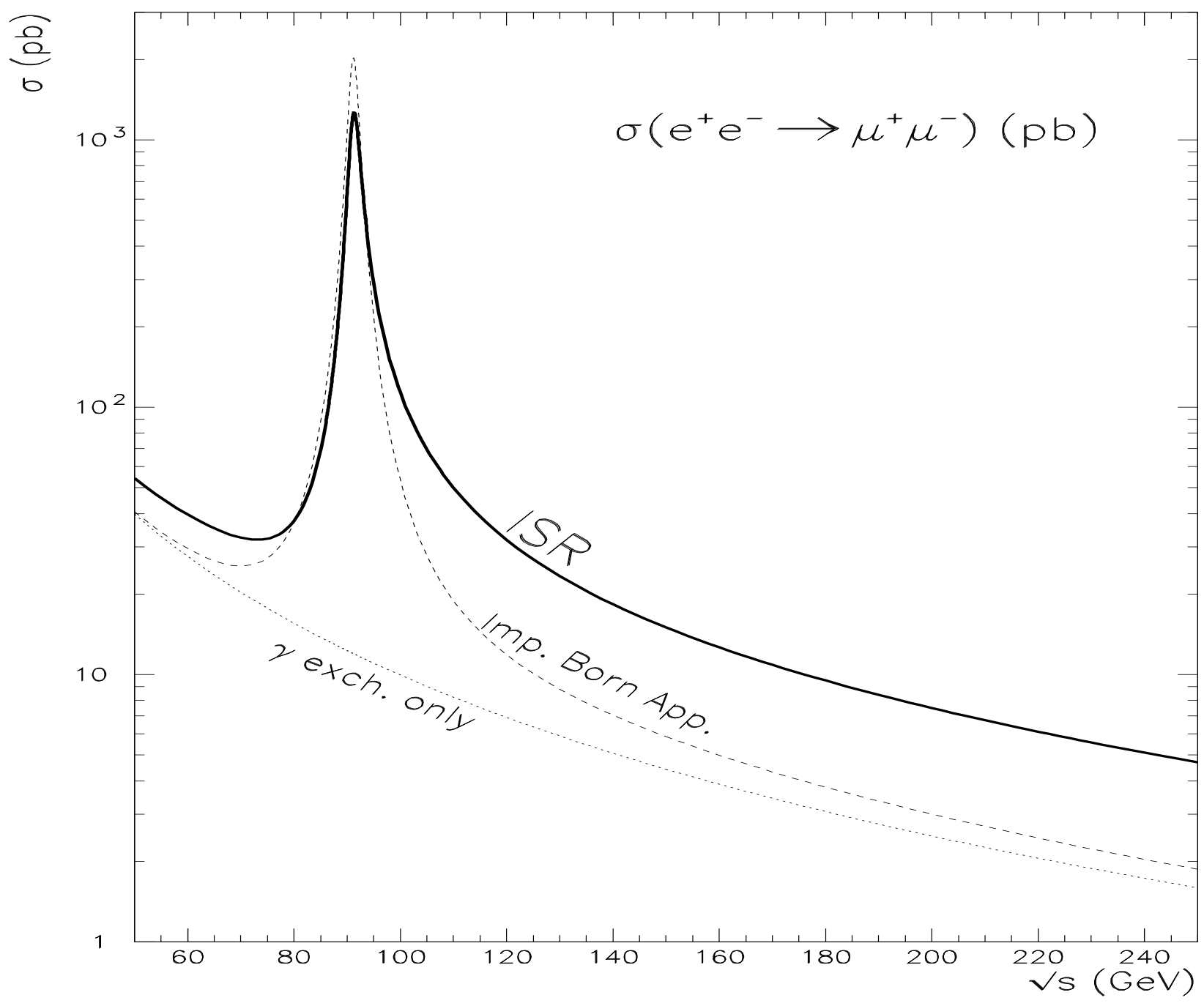

Figure 2: *The* $e^+e^- \ \to \mu^+\mu^-$ *cross section before and after the ISR convolution. ISR is included according to* [1].

## 2.2 Radiative corrections and status of tuned comparisons

Although there have been many exhaustive studies of two-fermion final states and many programs have been successfully tested, the comparisons among these programs have been performed and optimized for energies around the $Z$ peak. For a very recent state-of-the-art investigation see [4], where the main emphasis was put on the expected theoretical accuracy, assessed by comparing different codes with different implementations of the radiative corrections. However, as for the case of ISR pointed at earlier, a few characteristics of the two-fermion cross-sections are modified and new aspects appear when going to higher energies.
In order to address the issue of the status and the perspectives of tuned comparisons for $e^+e^- \to \bar{f}f$ processes at LEP2 energies, it is worth reviewing the various building blocks for computing observables related to the process $e^+e^- \to \bar{f}f$. We have:

*Here the $WW$ events add as "backgrounds" that necessitate extra cuts.

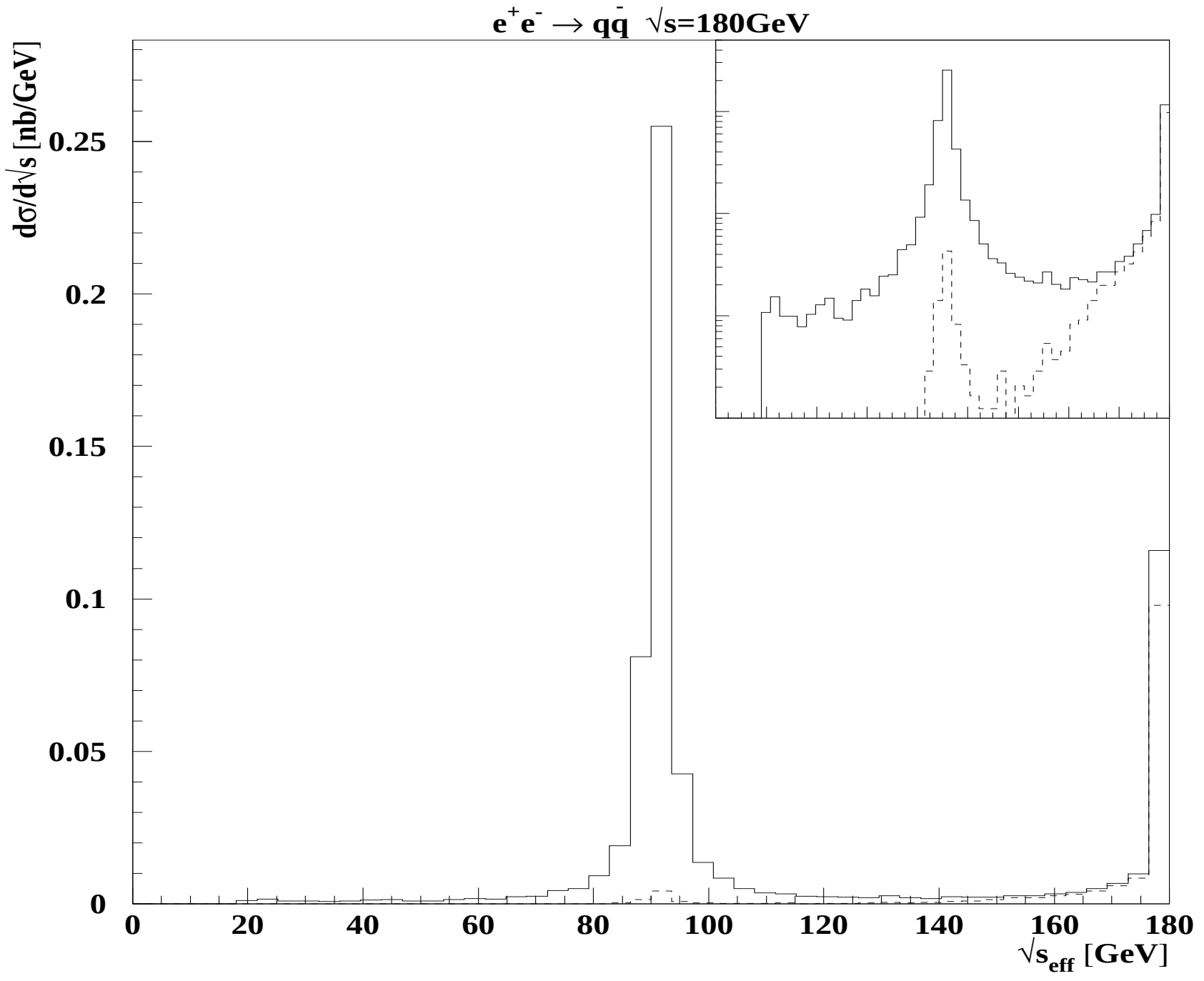

Figure 3: *The invariant mass distribution of the hadrons at a centre-of-mass energy of 180GeV before (solid) and after (dashed) cuts. The cuts reject an event if an isolated high energy photon is seen in the detector; if not the acollinearity of the two jets has to be less than* $20^0$ *and the observed invariant mass larger than* $0.4\sqrt{s}$. *The inlet is a blow-up (logarithmic scale) showing what remains of the Z return events after cuts.*

a) Pure electroweak corrections for the (kernel) deconvoluted distributions, including weak boxes ($WW$ and $ZZ$ internal lines). The latter were neglected at LEP1 energies since their relative contribution was of order $10^{-4}$.
b) Final state (FS) QED and QCD corrections.
c) Initial state (IS) QED radiation.
d) IS lepton- and quark-pair production (PP).
e) Initial-final (IF) QED interference.

The result of the implementation of each block is to be compared between different codes before a global comparison, which incorporates all the parts, is made. This not only avoids eventual accidental cancellations, but also brings out the relative contribution of the various "ingredients" entering in the totally convoluted "realistic observables". Within the study group, issue a) (deconvoluted observables) has been investigated by comparing the results of three codes based on different approaches for the implementation of the kernel: TOPAZ0 [5] $^\dagger$, WOH [7] and ZFITTER [8]. TOPAZ0 results have been computed in the 't Hooft-Feynman gauge, $\xi = 1$, and within the $\overline{MS}$ scheme, WOH has also $\xi = 1$ but on shell (OS) renormalization scheme

$^\dagger$Note that TOPAZ0 has been particularly designed to run around the $Z$ resonance and it is not optimized for much higher energies. For the LEP2 study, TOPAZ0 has been modified by upgrading the radiator function according to ref. [6] and including the contribution from weak boxes.

(RS) and, finally, ZFITTER works in the unitary gauge and in the OS RS. All three codes have adopted the input parameters:

$$\begin{aligned} M_Z &= 91.1884\,\text{GeV}, \qquad m_t = 175\,\text{GeV}, \\ m_H &= 300\,\text{GeV}, \quad \alpha_s(M_Z) = 0.123, \quad \alpha^{-1}(M_Z) = 128.896 \end{aligned} \tag{1}$$

A complete comparison incorporating all a)-e) (realistic observables) has been restricted to TOPAZ0 (T) *vs* ZFITTER (Z) and covered a sample of six energies: the $Z$ mass (in order to establish a link with the LEP1 calculations) and six LEP 2 energies, *i.e.* $140, 150, 161, 175, 190$, 205 GeV.

Because of the critical issue of the hard radiation at LEP2 energies and since both (T) and (Z) now apply the same QED radiator function for the total cross section, issue c) has been independently investigated by the Pavia group [9].

**• Pure Electroweak Corrections: effective couplings and the box problem**
The genuine electroweak corrections are by far the most interesting aspect of the two-fermion observables. Indirect virtual effects of new physics can also mimic these corrections. Hence, one needs to verify whether the strategies and approximations applied at the $Z$ peak are still at work. For example, a question related to the actual implementation of higher-order corrections is connected with the attempt of parametrizing physical observables in terms of 'running' effective couplings. This language of effective couplings, which has been so successful at LEP1, is deeply related to some factorization scheme that must be rediscussed at higher energies (for instance, weak boxes were neglected at LEP1). This language reduces the computational complexity, and does not introduce any $t$-dependence in the amplitudes, leading to a most useful and successful parametrization in terms of effective ($s$-dependent) vector and axial-vector couplings. Unfortunately, at LEP2 energies one expects the boxes to start resonating due to the $WW$ (and, to a lesser extent, ZZ) thresholds. Moreover, as can be inferred from a cut across these boxes which reveals the $e^+e^- \to W^+W^-$ $t$-channel exchange, the box contribution is not gauge invariant; in the same way that the $t$-channel for $e^+e^- \to W^+W^-$ is not unitary.
To quantify the effect of boxes one should first address some theoretical considerations about gauge invariance and give a procedure for isolating the effect of the weak boxes. It is well known that only a proper arrangement of the radiative corrections to $e^+e^- \to \bar{f}f$, including all contributions up to a given order, is gauge invariant. Every procedure designed for subtracting some part from the whole answer, for instance deconvolution of QED radiation, must respect gauge invariance. Formally, one writes the amplitude in terms of full 1PI vector-boson self-energies, initial(final) vertex corrections and multiparticle exchange diagrams. Next, the complex pole is derived in terms of the bare Lagrangian and, after a Laurent expansion, we end up with the pole, the residue at the pole and the *non-resonating background* (that encapsulates the $t$-dependence of the two-fermion amplitude), each of which is separately gauge invariant. It turns out that at LEP2 energies the *non-resonating background* (to which the boxes contribute) is not negligible. This is an unambiguous manifestation of the importance of the boxes.
Instead of using the complex pole formalism, which is difficult to implement in the codes, the effect of the boxes in the comparison has been handled by agreeing on a procedure for "extracting" the $WW$ exchange box diagram. Schematically, this diagram is denoted by $B_{WW}(\xi)$ as computed in a general $R_\xi$ gauge. It may be split, non-uniquely though, according to

$$B_{_{WW}}(\xi) = B_{_{WW}}(1) + \left(\xi^2 - 1\right) \Delta(\xi) = B_{_{WW}}(\xi_0) + \bar{\Delta}(\xi, \xi_0) \tag{2}$$

When working in the $R_\xi$ gauge, we can incorporate $\bar{\Delta}$ into the rest of the amplitude, which is $\xi$-dependent, and compute explicitly the $WW$ box diagram in any $\xi_0$ gauge. This approach is gauge invariant but not unique, especially when different procedures are adopted like keeping the weak boxes outside the QED convolution or performing re-summations. At this point we can adopt two different strategies. On the one hand, one can use the ZFITTER prescription of including the weak boxes into the form-factors. These then become explicit functions of the scattering angle. On the other hand, for comparison purposes, a proposal for "de-boxization" has been made. As one presents results for QED deconvoluted quantities, we could also subtract weak boxes from the data with few simple rules:

*i* It was agreed to substract $B_{_{WW}}(\xi = 1)$. At LEP1 this contribution can be neglected, its relative effect being of order $10^{-4}$.

*ii* those who work in the $\xi = 1$ gauge stop here,

*iii* those who work in any $\xi_0$ gauge compensate the rest of the amplitude with $\bar{\Delta}(\xi_0, 1)$.

The effect of weak boxes (as defined above) is studied on the deconvoluted observables $O_{dec}$ (*i.e.* before the inclusion of any IS and FS radiation) through the quantity $\delta_B$:

$$\delta_B = \frac{\sigma_{dec}}{\sigma_0} - 1 \qquad \text{where} \qquad \sigma_{dec} = \sigma_0 + \sigma_{box} \tag{3}$$

and $\sigma_0$ is the corrected cross section but without the inclusion of boxes.
First, for $e^+e^- \to \mu\mu, \bar{d}d, \bar{u}u$ TOPAZ0 and ZFITTER are found to agree extremely well from $\sqrt{s} = M_{_Z}$ up to LEP2 energies, including the region around the $WW$ threshold. The relative discrepancy is well below the per-mil level for both $\mu$ and $u$ and at worst 1.3 per-mil for $d$. There is some minor (in view of the expected experimental accuracy) disagreement with WOH($W$): $|\delta(W-T)|_\mu < 0.3\%$ , $|\delta(W-T)|_u < 0.5\%$ , $|\delta(W-T)|_d < 0.7\%$.
An important result, already pointed at, is that the effect of weak boxes is not negligible (a few per-cent in terms of $\delta_B$) especially around the $WW$ threshold and at the highest LEP2 energies. For instance, at 205 GeV, $\delta_B$ for the $\mu, d, u$ channels is $-1.1\%(-1.2\%)$, $-2.2\%(-2.6\%)$, $-3.4\%(-4.1\%)$ for TOPAZ0/ZFITTER(WOH). The results for the other energies are displayed in Table 1. This table also shows the effect of the boxes for $\sigma_{b\bar{b}}$ and $\sigma_{hadrons}$. For these two observables the comparison only involves TOPAZ0 and ZFITTER.

Comparisons of the results for $\sigma_{b\bar{b}}$ reveals a discrepancy between TOPAZ0 and ZFITTER which attains $\sim 2\%$ at $\sqrt{s} = 161 GeV$ and 205GeV while the agreement for $d$ is excellent. Looking in more detail, one sees that ZFITTER gives almost exactly the same $\delta_B$ for both $d$ and $b$. It should be remarked that, in ZFITTER (but not TOPAZ0) the top mass is neglected in the boxes. The TOPAZ0 results suggest that the inclusion of the mass decreases the relative effect of the box, as one would naively expect. This disagreement is in fact another indication of the special role played by the $b$ observables, a result reminiscent of the $Z \to b\bar{b}$ at LEP1 and the top connection. Note, however, that the $b$-box result does not have the same conceptual importance as the $Zb\bar{b}$ vertex in the sense of the non-decoupling of the heavy top (or equivalently the contribution of the Goldstone Bosons). Anyhow, this disagreement largely gets diluted and disappears when considering the total hadronic cross section. For the hadrons, the largest difference in $\delta_B$ (about 0.4%) shows up around the $WW$ threshold where the effect of the boxes is about 1%. The contribution of the boxes to $\sigma_h$ is larger at 190GeV and 205GeV, reaching about 2%.
Concluding on the effect of boxes and the comparison between the genuine weak corrections, we mention that the effect of boxes on the forward-backward asymmetry for muons is well below 0.01 and that TOPAZ0 and ZFITTER have been checked to agree perfectly for this observable.

| $E_{CM}$ (GeV) | ZFITTER | | | | | TOPAZ0 | | | | |
|---|---|---|---|---|---|---|---|---|---|---|
| | $\sigma^{\mu}$ | $\sigma^{u}$ | $\sigma^{d}$ | $\sigma^{b}$ | $\sigma^{h}$ | $\sigma^{\mu}$ | $\sigma^{u}$ | $\sigma^{d}$ | $\sigma^{b}$ | $\sigma^{h}$ |
| $M_Z$ | +0.00 | -0.01 | 0.00 | 0.00 | +0.00 | +0.00 | +0.00 | -0.01 | 0.00 | 0.00 |
| 100 | -0.59 | -3.56 | -0.50 | -0.51 | -1.59 | -0.59 | -3.33 | -0.40 | -0.51 | -1.41 |
| 140 | +0.09 | -11.09 | +1.02 | +0.94 | -4.71 | +0.07 | -10.69 | +0.84 | +0.00 | -4.66 |
| 150 | +1.79 | -10.00 | +4.32 | +4.36 | -2.59 | +1.81 | -9.58 | +4.09 | +0.51 | -3.10 |
| 161 | +11.75 | +3.52 | +24.63 | +24.90 | +14.16 | +11.81 | +3.37 | +23.38 | +3.86 | +10.17 |
| 175 | +2.02 | -12.64 | +5.10 | +5.14 | -3.94 | +1.84 | -12.37 | +4.67 | +0.80 | -4.29 |
| 190 | -5.29 | -25.25 | -10.48 | -10.60 | -18.12 | -5.58 | -24.69 | -10.25 | -2.30 | -16.45 |
| 205 | -10.53 | -34.89 | -22.38 | -22.56 | -28.95 | -10.95 | -34.11 | -21.66 | -5.18 | -25.60 |

Table 1: *The effect (in per-mil) of Weak Boxes, $\delta_B$, on $\sigma^{\mu}$, $\sigma^{u}$, $\sigma^{d}$, $\sigma^{b}$ and $\sigma^{h}$ before convolution.*

Once one has subtracted the effect of the boxes, the remaining building block of the genuine electroweak corrections are essentially those one has dealt with at length at LEP1 (apart from the fact that these are now evaluated at $k^2 \neq M_Z^2$). It is then worth inquiring about what one can learn from these "properly defined" observables that one has not from LEP1. In fact, one could further subtract the $k^2 = M_Z^2$ part and express the LEP2 observables in terms of the corresponding LEP1 quantities as suggested in [10]. The LEP1 observables were powerful enough in the sense that heavy particles, like the top, did not decouple, therefore allowing to put stringent limits on (or even ruling out) models beyond the $\mathcal{SM}$ . Unfortunately, after isolating the LEP1 observables, the remaining $k^2$ functions do not show much sensitivity to heavy particles, unless one is not far from their threshold. A most interesting topic concerning the $k^2$ dependence is the extraction of the running of $\alpha_{em}$. If this could be done unambiguously, in a gauge-invariant way, one might hope to measure the non-Abelian contribution to the running that exhibits anti-screening and which, at high-energy, slows down the growing of $\alpha_{em}$. It has been suggested to exploit the pinch technique [11], but more work related to this important issue is still needed.

**• ISR: Pure QED radiation**

We have already stressed the qualitative difference between initial state radiation at LEP1 *vs* LEP2 energies. Because of this important difference and the overwhelming effect of this "correction" (see Fig. 1 and Fig. 2), it is crucial to reassess the implementations of the ISR and then see how the convoluted "realistic" observables compare in different codes. This is most conveniently done by convoluting the weakly-corrected cross section $\sigma_{dec}$ (see Eq. 3) with a radiator function, $G(x = s'/s)$, that encapsulates the results of the QED corrections (virtual corrections and real radiation)

$$\sigma(s) \;=\; \int_{x_{cut}}^{1} dx \; \sigma_{dec}(s') \; G(x) \tag{4}$$

For a fully extrapolated set-up, $x_{cut} = 4m_f^2/s$. To cut the "Z-return" at LEP2, one may take $x_{cut} > 0.5$.

Clearly the $\mathcal{O}(\alpha)$ result in $G(x)$ is not sufficient. The $\mathcal{O}(\alpha^2)$ has been computed exactly[6] while one can resum (at least) the soft photons to all orders. This resummation is important especially for LEP1, and introduces an "exponentiation scheme ambiguity". A typical scheme, or parametrization for $G(x)$, after soft-photon resummation, is

$$G(x) = \beta\,(1-x)^{\beta-1}\delta_{S+V} + \delta_H(x), \qquad \beta = 2\,\frac{\alpha}{\pi}\,(L-1) \qquad L = \log\frac{s}{m_e^2} \tag{5}$$

where $\delta_{S+V}$ can be associated to the virtual and soft (bremsstrahlung) corrections while the additive $\delta_H$ is due to the hard-photon radiation (added linearly here). The large corrections are due to the "collinear logs" and formally one may write

$$\delta_{S+V} = \sum_{n=0}^{\infty}\left(\frac{\alpha}{\pi}\right)^n \sum_{i=0}^{n} \; S_{ni}L^i \qquad \delta_H(x) = \sum_{n=1}^{\infty}\left(\frac{\alpha}{\pi}\right)^n \sum_{i=0}^{n} \; H_{ni}(x)L^i \tag{6}$$

All schemes reproduce the leading logs, LL (i.e. $S_{nn}, H_{nn}$) up to some order $n$. For LEP1, $n = 2$ is sufficient. However, not all schemes reproduce even the <u>exact</u> $\mathcal{O}(\alpha^2)$ result. This difference is reflected essentially in the hard part and explains why schemes and codes (reproducing only the leading logs) that agree perfectly at LEP1 energies, no longer do so away from the resonance. Thus, TOPAZ0 has partly upgraded its radiator to reproduce the exact $\mathcal{O}(\alpha^2)$ result. Using the definitions set in ref. [12], both TOPAZ0 and ZFITTER use now a radiator function of type $G_A$ (i.e. of the same kind as in Eq. 5). $G_A$ includes the complete $\mathcal{O}(\alpha^2)$ corrections computed in [6].

One also expects that different parametrizations of the hard part affect mainly the inclusive cross section, while if a large $x_{cut}$ value is imposed, that cuts away the hard radiation, the difference between the two schemes gets substantially smaller. To quantify the impact of the hard radiation terms and make a comparison with the situation at LEP1, the effect of the order $\mathcal{O}(\alpha^3)$ LL versus the $\mathcal{O}(\alpha^2)$ (LL also) has been investigated in the case of $\sigma_{\mu\mu}$. The implementation of the LL is most easily and conveniently done within the framework of the QED structure-function approach. The cross section obtained with a standard (LEP1) additive structure function with up to second-order hard contributions ($\sigma_{Ad2}$) is compared against a non-singlet additive structure function inclusive of up to third-order hard-photon effects [13, 14] ($\sigma_{Ad3}$). Applied to LEP1 the relative contribution of the latter is below $10^{-4}$. Figure 4 clearly illustrates the points raised above [9]. While a stringent $x_{cut} > 0.5$ makes the higher-order effects completely negligible (below $10^{-4}$), for loose cuts $x_{cut} < 0.3$ (inclusive set-up) the effect lies in the range (0.1-0.4)% .

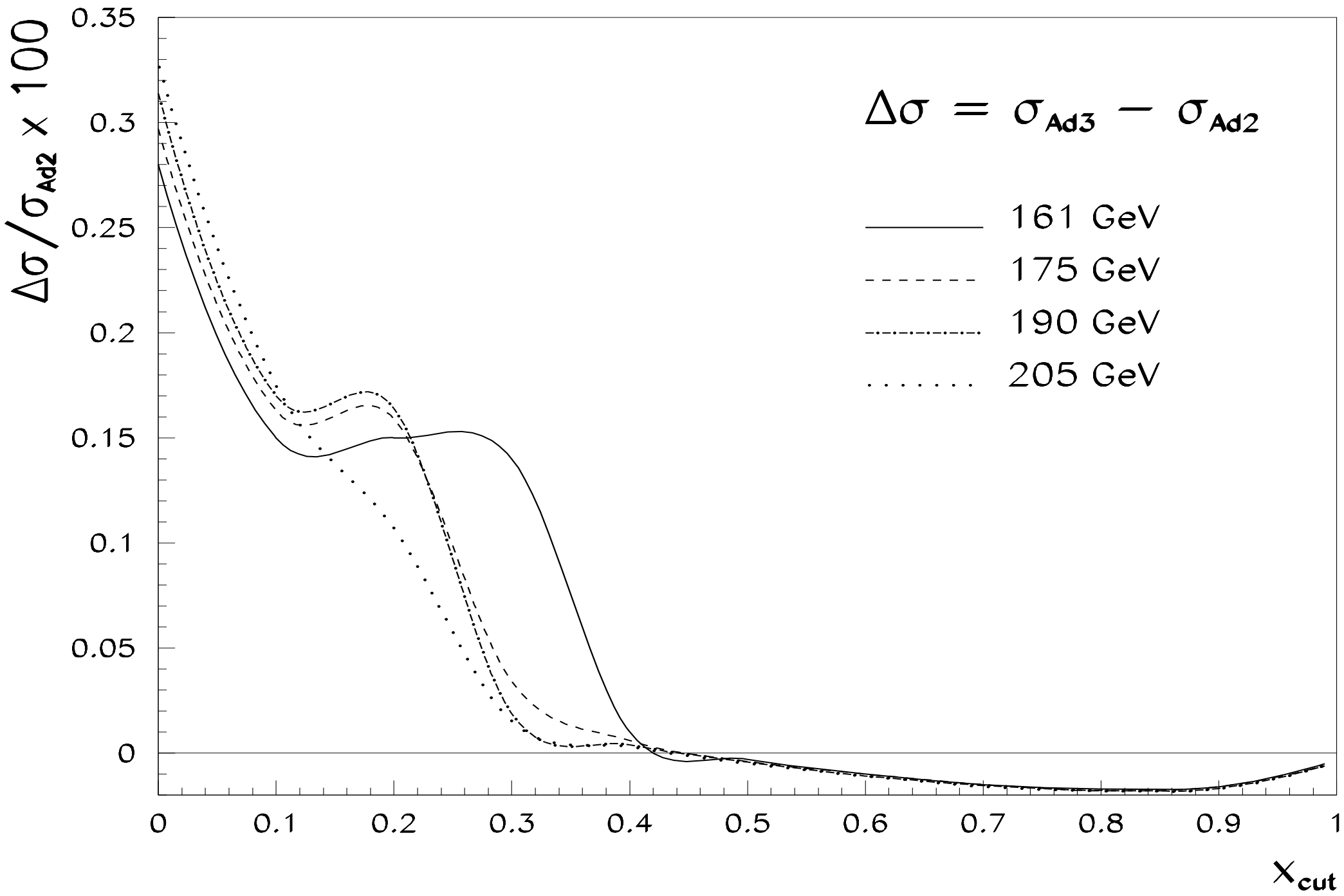

Figure 4: *The relative effect of the third-order leading log hard corrections on $\sigma_{\mu\mu}$ as a function of the cut on the invariant mass. Four LEP2 energies are considered.*

While ZFITTER and TOPAZ0 use the same radiator function for the total cross section, only ZFITTER implements the leading-log correction to the forward-backward asymmetry as given in [15]. The latter involves a "non-symmetric "(*i.e.* it would vanish upon angular integration) $\mathcal{O}(\alpha^2 L^2)$ contribution. This might be responsible for a tiny difference in $A_{FB}$. The comparison between TOPAZ0 and ZFITTER for the purely photonic convolution has been made with two values of $s'$: the first corresponding to an inclusive set-up with $s' > 0.015\,s$ and the second to a loose cut on $s'$: $s' > 0.5\,s$. In this first comparison about the effect of ISR, the contribution of the boxes discussed above is switched off. The results are displayed for the muon and hadronic cross sections, as well as the forward-backward asymmetry, in Table 2. For $\sigma_{\mu\mu}$ in the inclusive set-up, there is excellent agreement for all the energies considered with an almost constant relative deviation of about 3 per-mil. With a more stringent cut the agreement is further improved and almost reaches the accuracy achieved at LEP1. As for $A^{\mu}_{FB}$, the cut has little effect on the absolute deviation which never exceeds 0.006 and is thus very satisfactory. For the hadronic cross section, the worst deviation occurs at the $WW$ threshold, attaining 7 per-mil in the inclusive set-up, but is reduced by an order of magnitude when the $s'$ cut is applied. This table also shows the large reduction in the event sample when the stricter cuts are applied, hence getting rid of the Z-return, as we discussed above.

**• IS Pair Production (PP)**

A consistent treatment of initial state radiation at $\mathcal{O}(\alpha^2)$ should include the radiation of additional fermion pairs which also appear as a virtual correction at the two-loop level. Both TOPAZ0 and ZFITTER have used the available results at $\mathcal{O}(\alpha^2)$ of the KKKS formulation [16]. This takes into account soft pair radiation with all events radiated up to some energy $\Delta \ll \sqrt{s}$ and hard-pair radiation

| $E_{cm}$ (GeV) | $s' > 0.015\,s$ | | | $s' > 0.5\,s$ | | |
|---|---|---|---|---|---|---|
| | $\sigma^\mu(pb)$ | $\sigma^h(pb)$ | $A^\mu_{FB}$ | $\sigma^\mu(pb)$ | $\sigma^h(pb)$ | $A^\mu_{FB}$ |
| 91.1884 | 1477.8 | 30442 | $-0.12891\times10^{-2}$ | 1444.5 | 30320 | $-0.10435\times10^{-2}$ |
| | 1477.5 | 30444 | $-0.14298\times10^{-2}$ | 1445.8 | 30326 | $-0.11169\times10^{-2}$ |
| | 0.20 | -0.07 | 0.013 | -0.90 | -0.20 | 0.07 |
| 140 | 16.924 | 243.49 | 0.29892 | 7.4449 | 73.302 | 0.67141 |
| | 16.879 | 243.30 | 0.29787 | 7.4380 | 73.214 | 0.67705 |
| | 2.67 | 0.78 | 1.05 | 0.93 | 1.20 | -5.64 |
| 150 | 13.676 | 189.70 | 0.29608 | 5.9608 | 52.284 | 0.64787 |
| | 13.636 | 189.70 | 0.29659 | 5.9556 | 52.224 | 0.65244 |
| | 2.93 | 0.00 | -0.51 | 0.87 | 1.15 | -4.57 |
| 161 | 11.085 | 148.33 | 0.28855 | 4.8799 | 39.325 | 0.62009 |
| | 11.046 | 149.00 | 0.29436 | 4.8728 | 39.351 | 0.62396 |
| | 3.53 | -4.50 | -5.81 | 1.46 | -0.66 | -3.87 |
| 175 | 9.0385 | 118.10 | 0.28826 | 4.0189 | 30.521 | 0.59394 |
| | 9.0118 | 118.15 | 0.29194 | 4.0152 | 30.467 | 0.59749 |
| | 2.96 | -0.42 | -3.68 | 0.92 | 1.77 | -3.55 |
| 190 | 7.4763 | 95.922 | 0.28720 | 3.3516 | 24.394 | 0.57297 |
| | 7.4545 | 95.633 | 0.28982 | 3.3494 | 24.304 | 0.57635 |
| | 2.92 | 3.02 | -2.62 | 0.66 | 3.70 | -3.38 |
| 205 | 6.3184 | 80.021 | 0.28564 | 2.8479 | 20.119 | 0.55717 |
| | 6.2989 | 79.459 | 0.28811 | 2.8465 | 20.014 | 0.56043 |
| | 3.10 | 7.07 | -2.47 | 0.49 | 5.25 | -3.26 |

Table 2: *Comparing the results of the ISR convolution with boxes switched off. Two configurations are considered: $s' > 0.015\,s$ and $s' > 0.5\,s$. The first row is ZFITTER and the second one is TOPAZ0. The third row is the relative deviation (in per-mil) for the cross sections and ($10^3\times$) the absolute deviation for the forward-backward asymmetry.*

$$\begin{aligned}
\sigma_{pair} &= \sigma^{S+V}_{pair} + \sigma^{H}_{pair}\,, \\
\sigma^{S+V} &= \int_{4m^2}^{\Delta^2} dq^2 \int_{(\sqrt{s}-\Delta)^2}^{(\sqrt{s}-\sqrt{q^2})^2} ds'\, \frac{d^2\sigma^{4f}}{dq^2 ds'}\,, \\
\frac{d\sigma^H}{ds'} &= \int_{4m^2}^{s(1-\sqrt{s'/s})^2} dq^2\, \frac{d^2\sigma^{4f}}{dq^2 ds'}\,; \quad \sigma^H = \int_{sz_{min}}^{s(1-\Delta/\sqrt{s})^2} ds' \frac{d\sigma^H}{ds'}
\end{aligned} \tag{7}$$

The formula in Eq.7 involves two parameters $\Delta$ and $z_{min}$. The unnatural appearance of the infrared separator $\Delta$ makes questionable the exponentiation of soft pairs. In [17], an exponentiated result is given which is valid for leptons. No analogous treatment is available for hadrons, where the $\mathcal{O}(\alpha^2)$ result must be corrected for numerically when considering in addition IS photon radiation. No effort at all has been made so far in order to 'adapt' TOPAZ0 and ZFITTER for the treatment of radiated high-energy pairs. Both TOPAZ0 and ZFITTER do not exponentiate the "soft pairs" and $\sigma_{pair}$ is added linearly to the cross section. There is also the so-called $z_{min}$ problem [16]. IS PP has been successfully compared around the $Z$ resonance for various values of this parameter, and finally the default has been set to $z_{min} = 0.25$. This corresponds to an experimental selection of $Z$ decays where the invariant mass of the $Z$ products is at least 50% of the total and the soft–hard separator $\Delta$ has been fixed in the region where we see a plateau of stability. However, above the $WW$ threshold the four fermion channel becomes competitive and one must establish a clear separation between real four-fermion events (see the section on four-fermion production below) and IS pair-production corrections to two fermion events. It looks plausible to include into the corrections for two-fermion events only very soft leptonic and hadronic pairs, *i.e.* something like $z_{min} = 0.5, 0.6$ corresponding to 70.7% or 77.5% of $\sqrt{s}$ at $\sqrt{s} = 200\,\text{GeV}$. For the following comparisons, $z_{min} = 0.5$ has been chosen.

The effect of pure photonic and pair-production initial state radiation on $\sigma_{\mu\mu}$ is displayed in Table 3 in terms of the relative contribution of the "pairs", $\delta_p$. Although IS PP is very small (a few per-mil), we observe a much less satisfactory agreement between the two codes. With a $z_{min} = 0.5$ cut the agreement ZFITTER/TOPAZ0 is remarkable up to the maximum positive contribution, which happens to be around 100 GeV, after which there is a consistent difference of about $0.2 - 0.3\%$. Incidentally, for the inclusive $\sigma_{\mu\mu}$, this is of the same order as the discrepancy between the two codes when the IS PP is switched off (see Table 2), with the result that the two effects largely cancel. Inclusion of pair production at high energies requires more theoretical work.

| $E_{cm}$ (GeV) | ZFITTER | TOPAZ0 |
|---|---|---|
| 91.1884 | -2.57 | -2.50 |
| 100 | +4.41 | +4.96 |
| 140 | -2.95 | -0.59 |
| 150 | -3.43 | -0.80 |
| 161 | -3.58 | -0.90 |
| 175 | -3.90 | -0.97 |
| 190 | -4.10 | -1.03 |
| 205 | -4.27 | -1.07 |

Table 3: *The effect (in per-mil) of IS pair production, $\delta_p$, on $\sigma_{\mu\mu}$ with a cut $s' > 0.015\,s$ and $z_{min} = 0.5$. The first column is ZFITTER, the second is TOPAZ0.*

- **IF QED interference**

Although the formulations in TOPAZ0 and ZFITTER are totally independent, the IF QED interference has been tested successfully over the whole range of energies.

- **Global comparisons and realistic observables.**

For the global comparisons, all the ingredients listed above are included simultaneously. It should be noted that the weak boxes are added linearly to the cross section and are not convoluted with QED radiation. The outcome of this final overall confrontation are collected in Table 4 for the case of $s' > 0.5s$ (and $z_{min} = 0.5$). For this value of $s'$, the radiative $Z$ return at LEP2 would be effectively discarded, and the observables would be more sensitive to the high energy component of the kernel with the genuine electroweak corrections. For the muon cross section, there is a remarkable agreement between the two codes almost equalling the level of accuracy reached at LEP1. It is always below 0.2% at all energies. In fact, even when relaxing the $s'$ cut to switch to the inclusive cross section, the agreement is excellent and much better than the relative deviation observed in the case of the inclusion of pair production. As pointed out above, the "more-than-needed" accuracy is partly due to some cancellation. For $A^{\mu}_{FB}$, the relative deviation does not compete with the one observed at LEP1 energies, nonetheless it stays below the 0.01 mark. As mentioned earlier, part of the discrepancy may be attributed to the different inclusion of the pure QED ISR in TOPAZ0 and ZFITTER (the asymmetric $\mathcal{O}(\alpha^2)$ is not implemented in TOPAZ0). For the hadronic cross section, the agreement is quite satisfactory up to the $WW$ threshold. Beyond this energy, it somehow degrades and reaches even 0.8%. However, at these energies the box contributions (before convolution) were found to show some discrepancy (see Table 1) that goes in the same direction as the discrepancy revealed in the "realistic observable". Some of the deviation here should be attributed to the different treatment of the boxes for $\sigma_{b\bar{b}}$, and therefore one expects an improved agreement if the $b$ boxes are calculated with the same input parameters.
Let us also mention that KORALZ has also been upgraded for LEP2 energies. KORALZ[18] is a Monte-Carlo program for $e^+e^- \to 2f\ n\gamma\ (f = \mu, \tau, u, d, c, s, b, \nu)$ which includes YFS exclusive exponentiation of initial and final state bremstrahlung. Weak boxes are implemented. Full details may be found in [19].

| $E_{cm}$ (GeV) | $\sigma^{\mu}(pb)$ | $\sigma^{h}(pb)$ | $A^{\mu}_{FB}$ |
|---|---|---|---|
| 91.1884 | 1440.9 | 30243 | -0.06278$\times 10^{-2}$ |
| | 1442.2 | 30249 | -0.07646$\times 10^{-2}$ |
| | -0.90 | -0.20 | 0.14 |
| 100 | 110.47 | 2172.1 | 0.25148 |
| | 110.32 | 2169.5 | 0.25202 |
| | 1.36 | +1.20 | -0.54 |
| 140 | 7.5227 | 72.837 | 0.67444 |
| | 7.5147 | 73.108 | 0.68313 |
| | 1.06 | -3.71 | -8.69 |
| 150 | 6.0413 | 51.909 | 0.65176 |
| | 6.0347 | 52.157 | 0.65933 |
| | 1.09 | -4.75 | -7.57 |
| 161 | 4.9973 | 39.591 | 0.62250 |
| | 4.9883 | 39.748 | 0.62937 |
| | 1.80 | -3.95 | -6.87 |
| 175 | 4.0789 | 30.182 | 0.59910 |
| | 4.0737 | 30.363 | 0.60559 |
| | 1.28 | -5.96 | -6.49 |
| 190 | 3.3782 | 23.771 | 0.57940 |
| | 3.3747 | 23.950 | 0.58565 |
| | 1.04 | -7.47 | -6.25 |
| 205 | 2.8561 | 19.383 | 0.56462 |
| | 2.8535 | 19.548 | 0.57071 |
| | 0.91 | -8.44 | -6.09 |

Table 4: *Overall comparison with a cut* $s' > 0.5\,s$ *and* $z_{min} = 0.50$ *The first entry is ZFITTER, the second one is TOPAZ0. The third entry is the relative deviation (in per-mil) for the cross sections and* ($10^3\times$) *the absolute deviation for the forward-backward asymmetry.*

As a conclusion, fermion pair production is under control. The study has also revealed which particular points require further investigation, *i.e.* especially the treatment of IS pairs. An important fact is the confirmation of the importance of the box contribution for all the two-fermion channels, mainly at the $WW$ threshold and at the highest LEP2 energies. This should be kept in mind or compensated for when attempting to parametrize the two-fermion observables in terms of running effective couplings or $s$-dependent form factors. Another important aspect that needs a more detailed study is how different codes compare when realistic cuts (such as accolinearity cuts, cuts on the energy and scattering angle of the fermions) are applied on the fully dressed observables. A very preliminary investigation, restricted to the muon case, shows that the agreement between TOAPZ0 and ZFITTER somehow degrades when implementing an accolinearity cut. At the same time the integration error in TOPAZ0 is larger than what is at the Z peak. All this shows that more optimisation for LEP2 energies is needed, especially when introducing specific cuts.

# 3 Single-photon production

When studying fermion pair production the special case of neutrinos was not addressed, since this contributes an invisible cross section. At LEP1 [20], the latter can be inferred from the measurement of the lineshape, once it is assumed that all the visible modes are counted in $\Gamma_{e,\mu,\tau}$ and $\Gamma_h$. Another, less competitive, method at LEP1, is the measurement of the single photon yield from $e^+e^- \to \nu\bar{\nu}\gamma$. At LEP2 this technique is the only available way to reveal the production of stable neutrals. One may think of the supersymmetric neutralinos and sneutrinos or a fourth generation neutrino, to cite a few. For these heavy "beyond the $\mathcal{SM}$" neutrals, one needs to retain sufficient energy to produce them. Consequently, the associated radiation will tend to be softer than the typical photons coming from the $\mathcal{SM}$ radiative neutrino background. Actually, the latter are mostly very energetic and are easy to trigger on, since they are predominantly photons that recoil against real $Z^0$ decays to the 3 light neutrino pairs. Once again, we are dealing with the radiative $Z$ return, that produces very energetic photons. Thus the situation is much more promising than at LEP1.

## 3.1 Experimental requirements

Single-photon counting experiments at LEP1 have been rather delicate due to the essentially soft nature of the single photons. This necessitated low-trigger thresholds and high control of backgrounds in order to achieve sensitivity to the $Z^0$ invisible width [21]. In general, LEP1 experiments have required $E_\gamma > 1.5$ GeV and have restricted to the large polar angle region $|\cos(\vartheta)| < 0.7$. At LEP2, the photon energy spectrum from $e^+e^- \to \nu\bar{\nu}\gamma(\gamma)$ gives highly energetic photons which are easy to trigger on and can be measured well. The detectors are expected to function with similar performance as at LEP1. One relevant difference is that the minimum polar angle at which one can detect electromagnetic particles (veto angle), is likely to increase from about 25 mrad at LEP1 to about 33 mrad at LEP2 due to the installation of additional background shields to protect the tracking chambers from backgrounds produced at the higher beam energies. This will lead to more stringent cuts on the transverse momentum scaled to the beam energy, $x_T = p_T/E_b$, designed in order to kinematically eliminate backgrounds from, *e.g.*, radiative Bhabha scattering (where the two electrons are below the veto angle). Moreover, LEP2 physics studies involving ISR can use the forward acceptance more readily than at LEP1, due to the much higher energies involved. Details depend on backgrounds and requirements: counting or measuring. Based on current analyses (studying for example $e^+e^- \to \gamma\gamma$) acceptance in polar angle in the region $|\cos(\vartheta)| < 0.95$ should be achieved for photons with sufficient $p_T$. Using canonical cuts of $x_T = p_T/E_b > 0.05$ and $|\cos(\vartheta)| < 0.95$, leads to a cross-section of about 5 pb for $\sqrt{s} = 180$ GeV. So, there is potential for a 2% measurement of the inclusive cross-section per experiment for an integrated luminosity of 500 pb$^{-1}$, indicating that a theoretical precision below 1% (4 experiments) is desirable. Given the striking nature of such events and the favourable energy spectrum, it is likely that measurements will be statistics limited and not limited by experimental systematics. On the other hand, to achieve this accuracy a precise knowledge of the $\mathcal{SM}$ cross section is needed, also taking into account that some approximations used at LEP1 are no longer valid. Moreover, of particular interest to experimentalists is the inclusion in Monte Carlo codes of additional hard photons to the single photon, as such photons affect the acceptance when emitted at detectable polar angles.

## 3.2 Calculations for $e^+e^- \to \nu\bar{\nu}\gamma$: lowest-order and radiative corrections

Neutrino-pair production in association with a photon is not entirely due to $Z$ decays. For $\nu_e$ there are additional $W$-exchange diagrams where the photon can be an ISR or from "internal radiation" (involving the non-Abelian $WW\gamma$ vertex). These $W$-exchange diagrams are not negligible at all at LEP2 energies, contrary to LEP1 [20]. For instance, comparing the purely $s$-channel $\nu_\mu\bar{\nu}_\mu\gamma$ with $\nu_e\bar{\nu}_e\gamma$, there is about a factor 2 enhancement of the latter at $\sqrt{s} = 175$GeV, brought about by the $t$-channel. This applies for a visible photon with $E_\gamma > 10GeV$ and $|\cos\theta_{e\gamma}| < 0.9$. Therefore, some of the approximations that worked so well for LEP1 and allowed for an easy implementation of the higher-order corrections are no longer valid.
An excellent approximation at LEP1, the so-called PIA [22], is obtained by taking the $Z$ contribution complemented by the limit $M_W \to 0$ (and switching off the $WW\gamma$). This reproduces the exact result within 1%. Another equally good approximation convolutes the neutrino-pair cross section (with $M_W \to 0$) with a radiator function[23]. These same approximations overestimate the result of the full calculation by some 30%, already at $\sqrt{s} = 150GeV$[‡].
Because of the failure of these approximations as the energy increases, the implementation of the higher-order corrections for this three-body reaction requires a special attention. Full one-loop QED corrections have been computed [25, 26], while complete $\mathcal{O}(\alpha)$ weak corrections are presently known only for the "sub-process" $e^+e^- \to Z\gamma$ [27]. Higher-order QED corrections, necessary to match the experimental precision reached at LEP, are taken into account in the

[‡]This is obtained with $E_\gamma > 1GeV, |\cos\theta_{e\gamma}| < 0.966$, [24].

Monte Carlo [24, 28] and semi-analytical codes [23] used by LEP collaborations, through the QED structure-function approach (SF) or the YFS algorithm to implement multiphoton effects. For instance, within the SF method [29], the QED-corrected cross section can be written as [30]

$$\sigma(s) = \int dx_1\, dx_2\, dE_\gamma\, dc_\gamma\, D(x_1, s) D(x_2, s) \frac{d\sigma}{dE_\gamma dc_\gamma}, \tag{8}$$

where $d\sigma/dE_\gamma dc_\gamma$ is the exact spectrum of Ref. [25], the photon variables refer to the centre-of-mass frame after initial-state radiation, and $D(x,s)$ is the electron (positron) structure function. The explicit expression of $D(x,s)$, including soft multiphoton emission and hard collinear bremsstrahlung up to $\mathcal{O}(\alpha^2)$, can be found in [30].

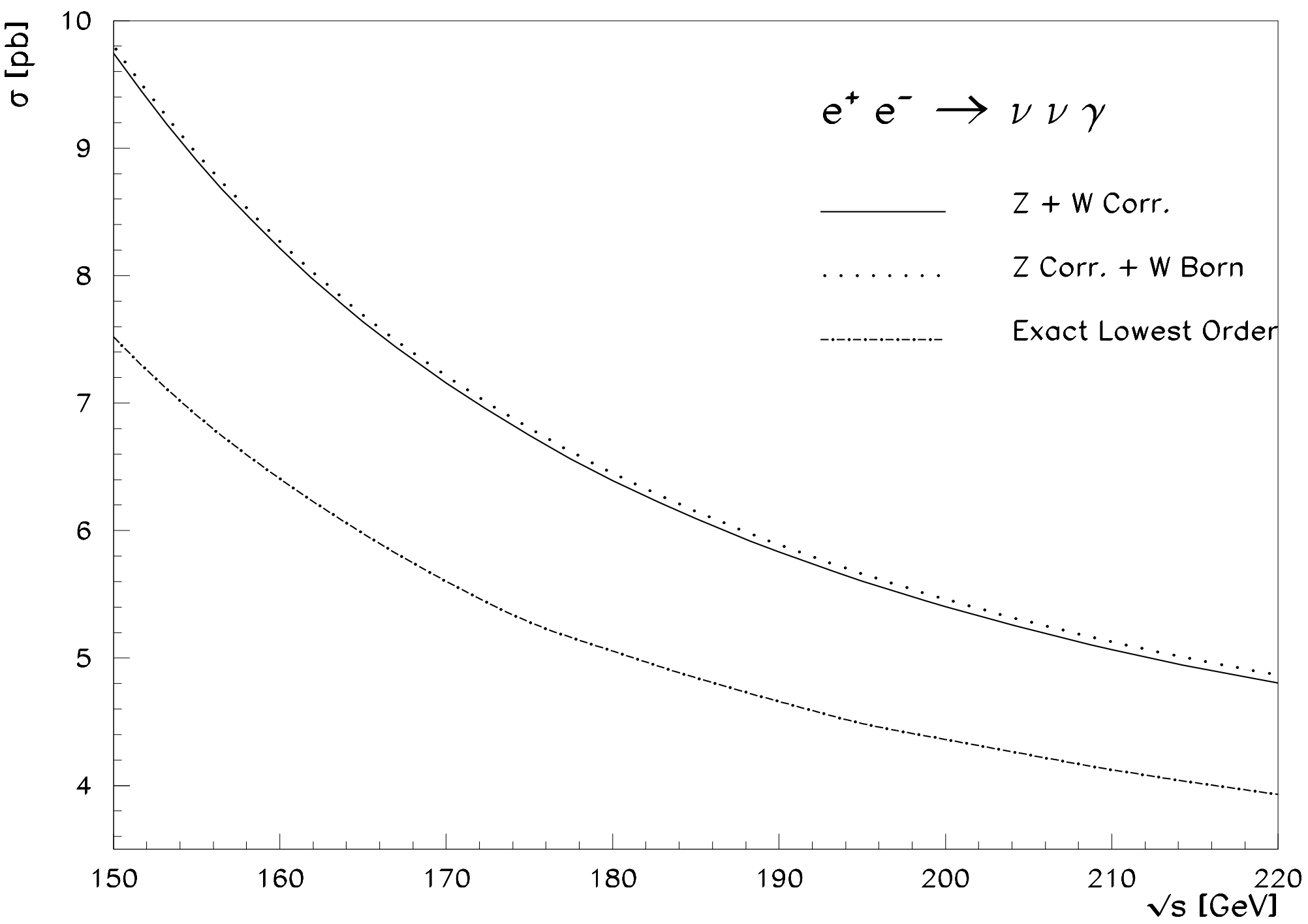

Figure 5: *The QED-corrected and the Born cross section for* $e^+e^- \to \nu\bar{\nu}\gamma$ *as a function of the centre-of-mass energy at LEP2. The approximations are detailed in the text. The cuts on the photon are* $E_\gamma^{min} = 1$ *GeV and* $\vartheta_\gamma^{min} = 20°$.

Fig. 5 shows the QED-corrected cross section of $e^+e^- \to \nu\bar{\nu}\gamma$ as a function of the centre-of-mass energy at LEP2, assuming the cuts $E_\gamma^{min} = 1$ GeV and $\vartheta_\gamma^{min} = 20°$. The dash-dotted line represents the lowest-order total cross section obtained after integrating the exact photon spectrum $d\sigma/dE_\gamma d\cos\vartheta_\gamma$ [25], the solid line is the result obtained according to the structure-function formulation of Ref. [30] (i.e. in the case of convolution of the full spectrum [25], including $Z$ and $W$ diagrams), and the dotted line, reported for the sake of comparison, shows the results obtained by simulating the approach of Ref. [24], namely correcting the $Z$ contribution only, and adding to this result the $W$-exchange diagrams at tree level[§]. Two considerations are in order. First, the QED-corrected cross section at LEP2 is higher than the Born one as a consequence of the $Z$ radiative return: the effect is to enhance, in this experimental set-up, the Born cross section by a factor of about 1.3. Secondly, the convolution of the full spectrum is in good agreement (within 1%) with the approach based on Ref. [24] because the QED-corrected cross section is largely dominated by the $Z$ radiative return, and the tree-level contribution of $W$ diagrams and $W$–$Z$ interference is almost flat over the full energy range spanning from LEP1 to LEP2. The agreement between the two calculations is within the expected experimental accuracy.

[§]Strictly speaking, in Ref. [24] the SF approach is applied to the complete $\mathcal{O}(\alpha)$ QED corrections to the $Z$ exchange contribution of $e^+e^- \to \nu\bar{\nu}\gamma$. In the comparison reported here, the SF is applied to the tree level

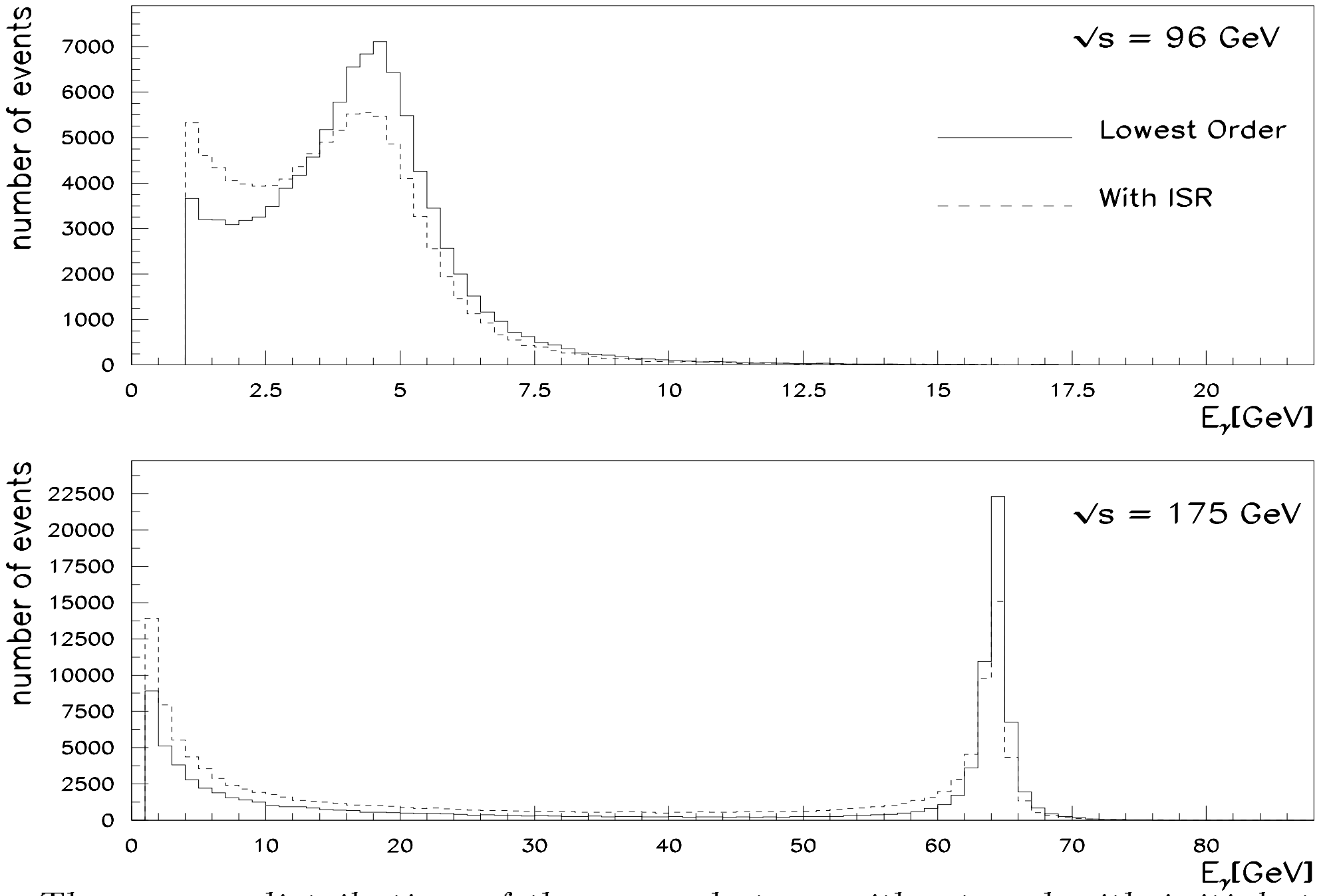

Figure 6: *The energy distribution of the seen photon, without and with initial-state QED corrections, for a LEP1 ($E_b = 48$ GeV) and a LEP2 ($E_b = 87.5$ GeV) energy. The cuts on the seen photon are the same as with the previous figure. The numbers of events integrated in the case with ISR and without are proportional to the corresponding integrated cross sections.*

The single-photon energy distribution is shown in Fig. 6 after including the higher order ISR. The results confirm the qualitative arguments given above concerning the LEP2 *vs* LEP1 comparison. Two peaks are clearly visible in the photon energy distribution, both at LEP1 and LEP2 energies: the higher one is located at the energy value of about $(1-M_Z^2/s)\sqrt{s}/2$, the lower one is due to $1/E_\gamma$ (soft photon) peaking behaviour. As can be seen, the main modifications introduced by initial-state radiation are to reduce the higher peak and to enhance the lower one. The most important conclusion is that at LEP2 even after taking into account additional radiation, there still is a prominent peak around the recoil hard photon, hence allowing for a better discrimination of the heavy neutrals and improving the LEP1 limit on the number of neutrinos via the radiative method; even if this will not match the super-precision of the line-shape method. Further simulations of single-photon distributions at LEP2 versus LEP1

---

$e^+e^- \to \nu\bar{\nu}\gamma$ cross section. Weak corrections are implemented through an improved Born approximation in both cases. It has been checked that both versions agree within 1% with the approach of convoluting the full spectrum.

energies obtained analyzing the events generated by the Monte Carlo of Ref. [31] are given and commented in Ref. [30]. All the above results have been produced by means of a new Monte Carlo event generator [31] developed for radiative neutrino counting measurements at LEP1/LEP2 and based on eq. (8).

### 3.3 Towards a single-photon library

During this Workshop, the problem of finding a general approach to the computation of the single-photon spectrum associated to any process of the kind $e^+e^- \to (invisible)$ has been addressed. In particular, possible approximations have been studied that, starting from the $e^+e^- \to (invisible)$ cross section, could allow to get the corresponding single-photon spectrum in a straightforward way. The Standard Model process $e^+e^- \to \nu\bar{\nu}\gamma$ can act as a benchmark for this purpose. Instead of the exact formula for the neutrino single-photon spectrum, one can use as a kernel in the convolution formula (8) an approximate factorized photonic spectrum given by

$$\frac{d\sigma^{approx}}{dx_\gamma dc_\gamma} = \sigma_0((1 - x_\gamma)\, s) H^{(\alpha)}(x_\gamma, c_\gamma; s), \tag{9}$$

where $\sigma_0$ is the total Standard Model cross section of $e^+e^- \to (Z, W) \to \nu\bar{\nu}$ and $H^{(\alpha)}(x_\gamma, c_\gamma; s)$ is the angular radiator proposed in Ref. [30] and derived from $\mathcal{O}(\alpha)$ $p_t$-dependent structure functions [32]. It describes the probability of radiating a photon with a given energy fraction $x_\gamma = E_\gamma/E_b$ at the angle $\vartheta_\gamma$ ($c_\gamma \equiv \cos\vartheta_\gamma$).

This approximation can be used as a basic tool to develop a library of single-photon events, including standard and non-standard (in particular SUSY) processes. Indeed, given as a kernel the total cross section corresponding to a process of the type $e^+e^- \to (invisible)$ objects, dressing it with the angular radiator $H^{(\alpha)}$, according to eq. (9), amounts to attaching a photon line on the external charged legs, including the "universal", factorized form of the photonic radiation. The above recipe has been checked against the exact Standard Model single-photon spectrum and found to be accurate at the level of a few per cent [30]. The same method has very recently been applied to the single-photon signature of the SUSY process $e^+e^- \to \chi\chi\gamma$ (for the most general gaugino/higgsino composition of neutralinos in the MSSM)[33]. Its cross section has been obtained by convoluting the cross section for the channel $e^+e^- \to \chi\chi$ with the radiator function and found to be very hard to disentangle from the neutrino background (see the Neutralino Section in the *New Particles* Report for some results on this channel).

## 4 Photon-pair production

Photon-pair production is essentially a pure QED process, that is not very sensitive to the genuine weak radiative corrections. Therefore, contrary to the single-photon production, there is no new phenomenon to take into account with respect to LEP1. One way to exploit this clean channel is to probe the indirect effects of alternative models such as the exchange of a heavy excited electron or a contact interaction. However, to conduct these tests it is essential to take into account the order $O(\alpha^3)$ QED corrections that could mimic new-physics effects. The corrected differential cross section may be written as:

$$\left(\frac{d\sigma}{d\Omega}\right)_{\alpha^3} = \left(\frac{\alpha^2}{s}\frac{1+cos^2\theta}{1-cos^2\theta}\right)(1 \; + \; \delta_{QED}) \tag{10}$$

where $\theta$ is the photon scattering angle with respect to the beam. $\delta_{QED}$ includes the virtual, soft and hard bremsstrahlung corrections [34]. This higher-order factor has been verified to be needed in order to reproduce the LEP1 data [35] as shown in Figure 7.

This correction will have to be included also at LEP2. However, one expects the sensitivity to the anomalous effects to be enhanced at LEP2, since the latter increase with energy, while the QED cross sections falls. For instance, the effect of an excited heavy electron that may be parameterized by a scale $\Lambda_\pm$ (depending on the chirality of the coupling) [36] or a general dimension-6 contact interaction with a scale $\Lambda$ [12, 37] modify the differential cross section according to

$$(d\sigma/d\Omega) = \; (d\sigma/d\Omega)_{\rm QED}\,(1 \; + \; \delta_{\rm new}) \tag{11}$$

where $\delta_{\rm new} \;\cong\; \pm s^2/2 \;\left(1/\Lambda^4_\pm\right)\; (1-\cos^2\theta)$ for the excited electron assumption and with an analogous expression for the contact interaction. A comparison of the measured and QED predicted differential cross sections, including the deviation, are reproduced from the L3 experiment in Figure 8. At LEP2, with an integrated luminosity of about 66 $\rm pb^{-1}$, the lower limit on the scale of the contact term $\Lambda$ is expected to increase from 600 to 800 GeV, while that describing the excited electron, $\Lambda_+$ and $\Lambda_-$, will go up to 200 GeV. These limits scale as the 1/4 power of the integrated luminosity.

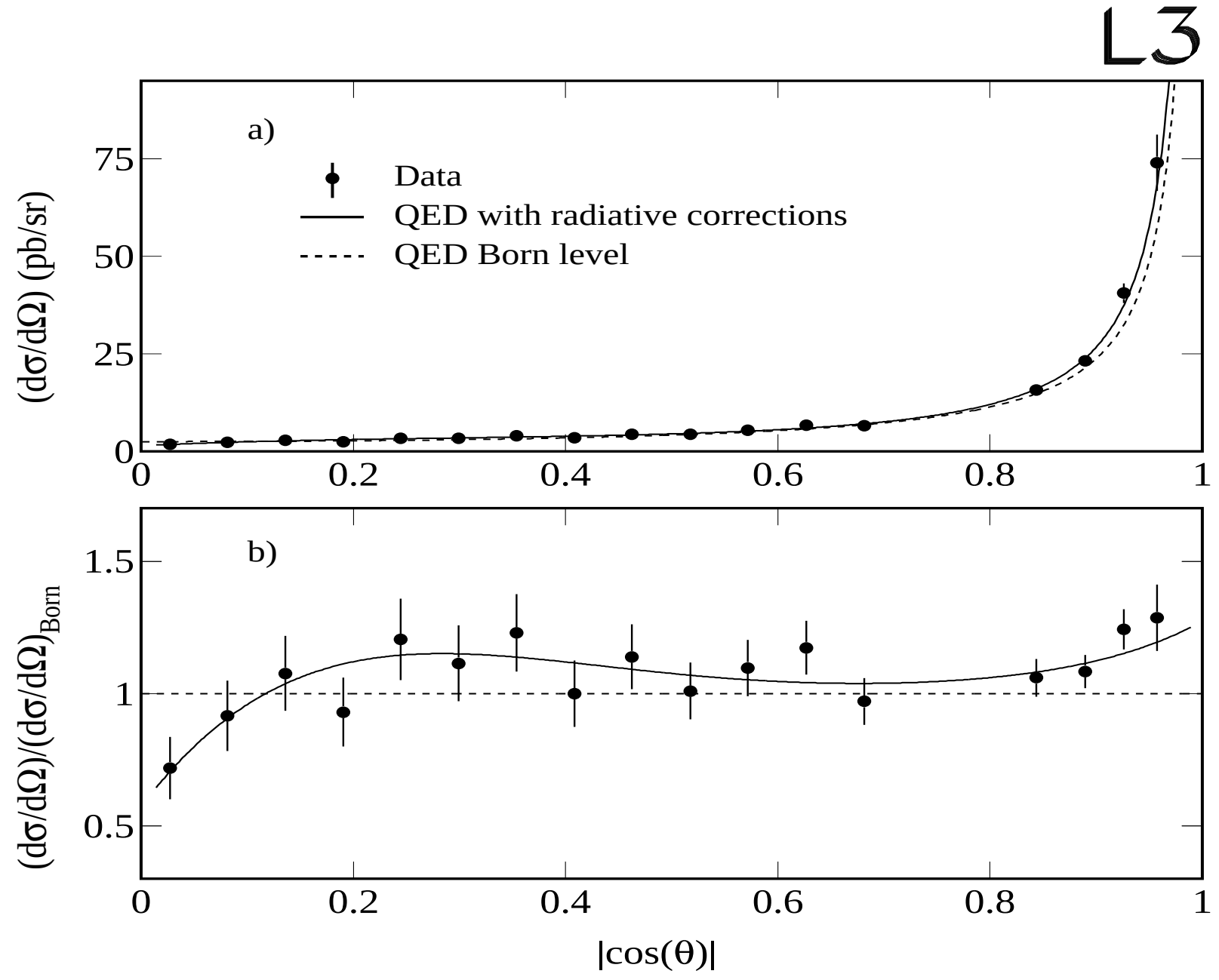

Figure 7: (a) *shows the comparison of the measured differential cross section with the QED prediction for the process* $e^+e^- \to \gamma\gamma(\gamma)$ *as a function of* $|\cos\theta|$. (b) *shows the same cross sections normalized to the QED Born level prediction. The comparison leads to a* $\chi^2$ *= 0.53/dof.*

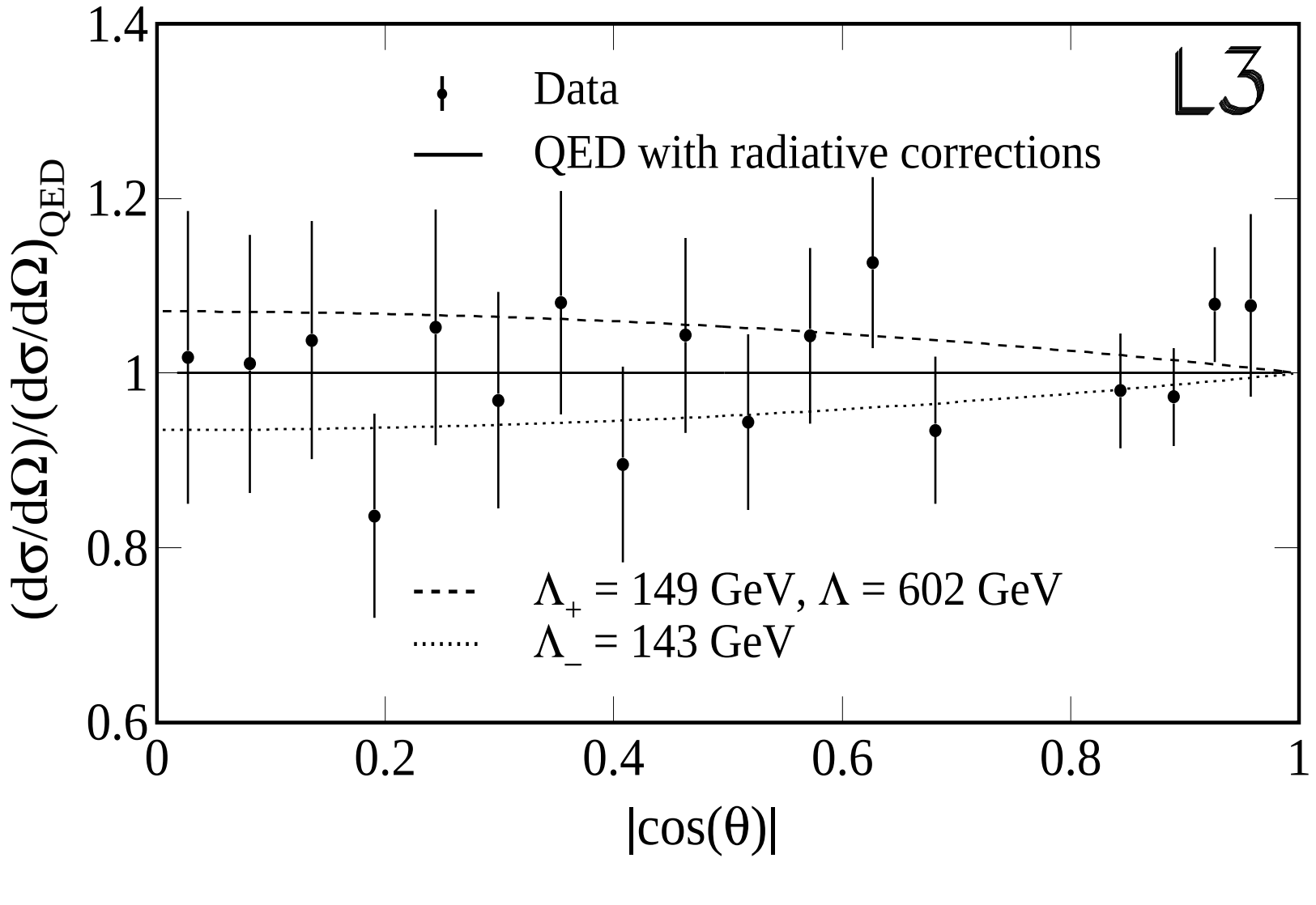

Figure 8: *Comparison of the measured differential cross section with the QED predictions including the deviations for the parameter values shown in the figure, as a function of* $|\cos\theta|$. *The cross sections are normalized to the radiatively corrected QED cross section. The functional effect of* $\Lambda_+$ *and* $\Lambda$ *is the same.*

# 5 Four-Fermion Processes

## 5.1 Classes of Feynman diagrams

At LEP2 centre-of-mass energies, four-fermion final states are produced with large cross sections. These are not only due to real $WW$ and $ZZ$ pair production with subsequent decays $W \to \bar{f}f'$ and $Z \to \bar{f}f$, but arise from several production mechanisms, each giving sizeable contributions to the four-fermion cross section in specific configurations of the final-particle phase space. In Fig. 9, all the possible classes of four-fermion production diagrams are shown. The largest total cross sections arise from the *multiperipheral* diagrams. Here, two quasi-real photons are exchanged in the $t$-channel, giving rise to forward (and undetected) electrons/positrons plus a $\bar{f}f$ pair with a non-resonant structure (the so-called "two-photon" processes). For instance, one has $\sigma(e^+e^- \to e^+e^- \tau^+\tau^-) \sim 10^2$ pb for $M_{\tau\tau} > 10$GeV. On the other hand, although interesting for QCD studies (see the $\gamma\gamma$ *Physics* report) and as a main background for missing energy/momentum events (see the *New Particles Physics* report), these classes of processes do not sizeably contribute to final states that are of interest for the studies of $W, Z$ and Higgs boson production. In the latter case, the main contributions come from double-resonant diagrams (*conversion* and *nonabelian-annihilation* diagrams in Fig. 9). Also single-resonant processes

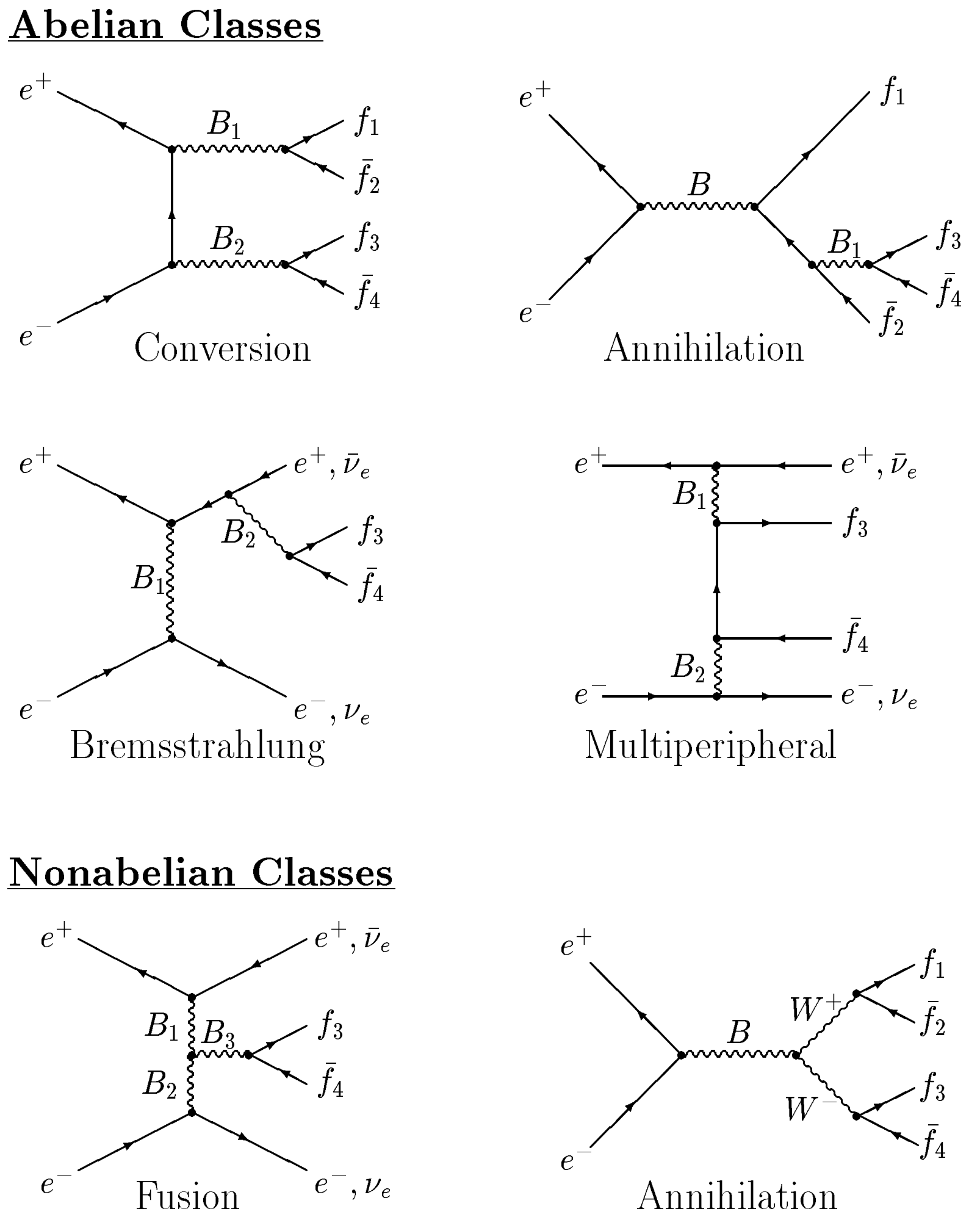

$(B\!=\!Z^0, \gamma;\;\; B_1, B_2, B_3\!=\!Z^0,\; \gamma,\; W^{\pm};\;\; + \, Higgs\ Graphs.)$

Figure 9: *Four-fermion production classes of diagrams.*

(proceeding through *abelian-annihilation, bremsstrahlung, fusion* and single-resonant *conversion* graphs) can give an important contribution to vector-boson physics, when the invariant mass constraint on one of the final fermion pairs is relaxed. A particular example is given by the single $W, Z$ production, $e^+e^- \;\to e\nu W \to e\nu f f'$ and $e^+e^- \;\to eeZ \to eeff$. In this case, most of the cross section is due to single-resonant *bremsstrahlung* and *fusion* diagrams, where an almost real photon is exchanged in the $t$-channel and one final electron escapes detection. In a sense, one could rename these channels as "three-(*visible*)fermion" processes.
Some aspects of four-fermion processes are studied elsewhere in this report. Here we concentrate essentially on providing analytical (or semi-analytical) approaches. A particular attention is given to total cross sections especially in the case of forward electrons. We will also list the cross sections for the entire list of the four-fermion processes when some canonical cuts are imposed, as given by some available codes on the market, thus complementing the studies of the *Events Generators for WW Physics* group.

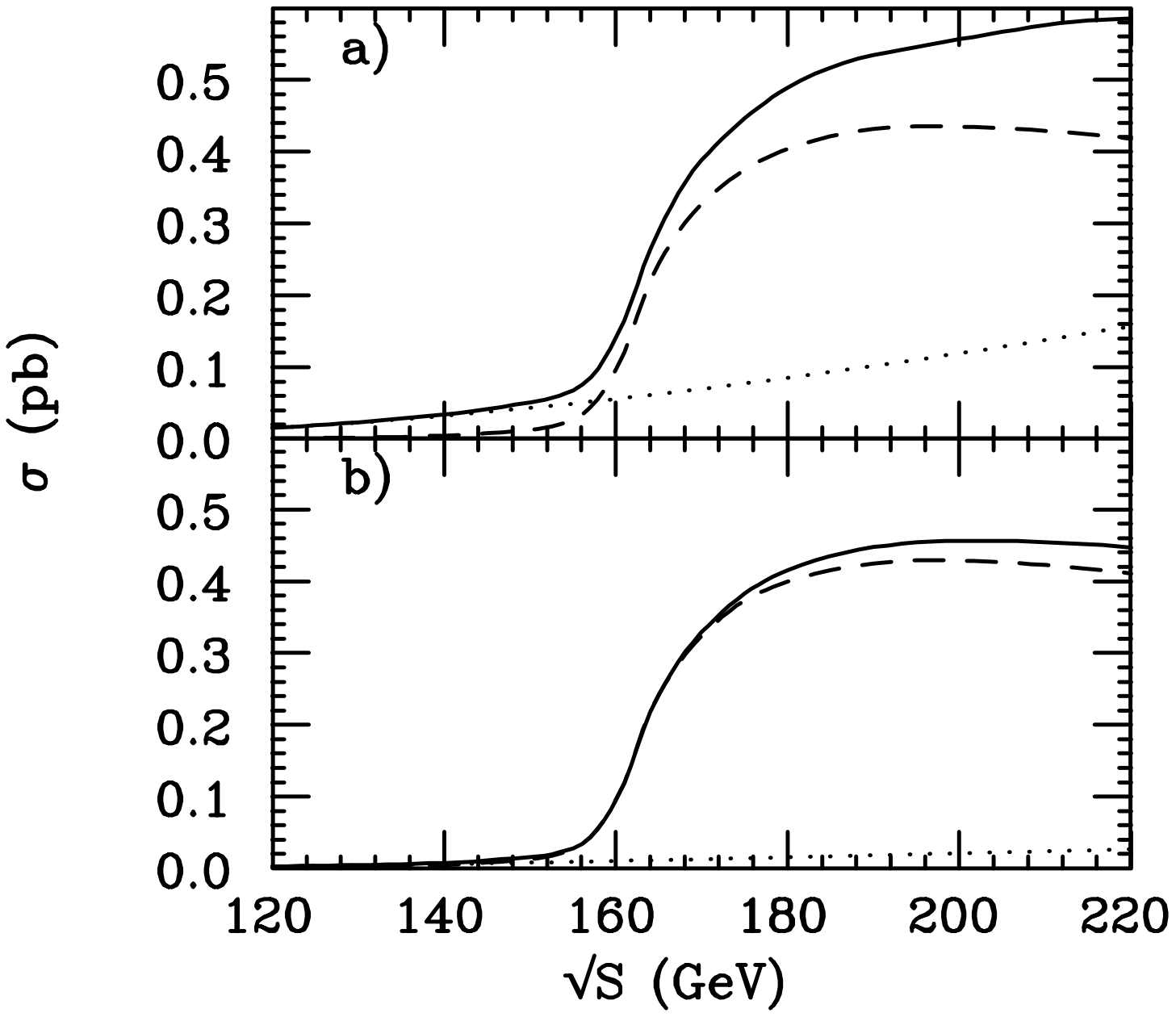

Figure 10: *Total cross-section for* $e^+e^- \to e^-\bar{\nu}u\bar{d}$*, all diagrams (solid), for, a), no cut on the final electron, and b), a cut on the electron angle with respect to the beams* $\theta_e > 8^o$*. Dashed and dotted lines show the double-resonant and t-channel contribution , respectively.*

### 5.2 Single-$W$ production

The cross section for single (on-shell $W$) production is shown in Fig. 1 and is dominated by the $t$-channel photon exchange. However, this is only one of the sub-processes that contributes to $e^+e^- \to e^-\bar{\nu}u\bar{d}$. Complete tree-level cross sections for the process $e^+e^- \to e^-\bar{\nu}u\bar{d}$ have been computed using the GRACE system [38] with the complete set of tree-level diagrams and taking into account all fermion-mass effects. This allows to integrate with no cuts over the forward-electron angle and exactly assesses the relative importance of double-resonant $W$ diagrams versus single-resonant $W$ and non-resonant diagrams [39]. In Fig. 10, after applying some realistic experimental cuts on the quark ( $E_{u,\bar{d}} > 1$GeV and angular separation from the beam $\theta_{u,\bar{d}} > 8^o$), the comparison of the total cross sections for all the diagrams (that is, 20 graphs) with the double-resonant subset (given by *conversion* plus *nonabelian annihilation* graphs, total of 3) and the $t$-channel subset (given by *bremsstrahlung, fusion* and *multiperipheral* graphs, total of 10) is shown for a) no cut on the final electron, and b) a cut on the electron angle with respect to both beams $\theta_e > 8^o$. The dominant contribution to the $t$-channel subset is given by the 4 diagrams where a photon is exchanged in the $t$-channel, with the $e^-$ scattered in the forward direction ¶. One can see that, below the $WW$ threshold, the single-resonant and non-resonant diagrams give a substantial contribution to the total cross section. At $\sqrt{s} = 190$GeV, their contribution is 4.4% of the total, while it increases at larger $\sqrt{s}$. On the other hand, imposing a cut on the forward electrons strongly depletes the $t$-channel contribution.

¶There is a very subtle problem with the implementation of the $W$ width. A naive "running" width leads to disastrous predictions, see [39]. A general discussion about the implementation of the $W$ width and gauge

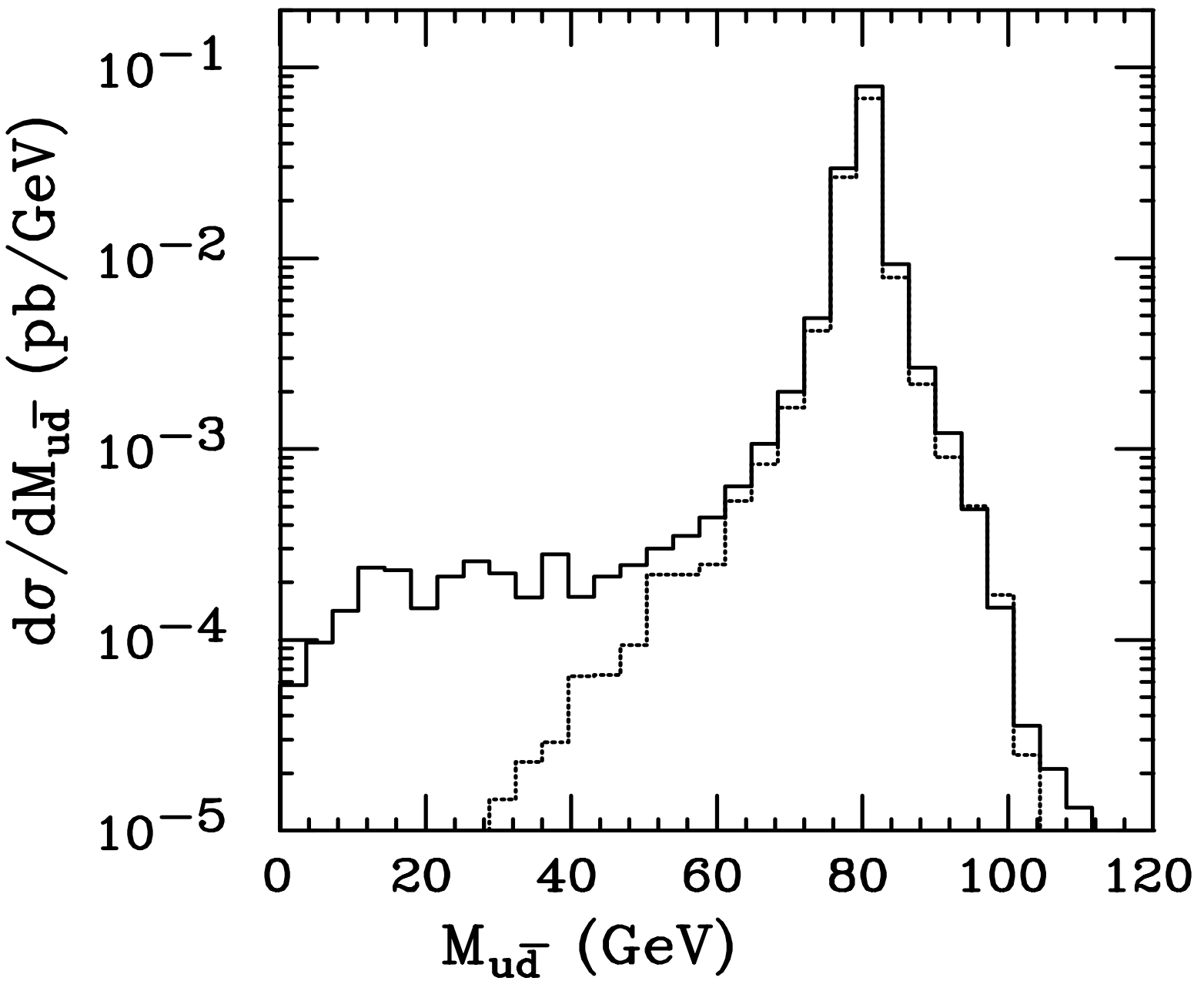

Figure 11: *Invariant $u\bar{d}$-mass distribution for $e^+e^- \to e^-\bar{\nu}u\bar{d}$ at $\sqrt{s} = 180$ GeV. The solid and dashed curves are, respectively, as for cases a) and b), of the previous figure.*

It is also interesting to compare the effect of the single-resonant and non-resonant diagrams on the quark-pair invariant mass distribution. Figure 11 shows how $t$-channel production can alter the $M_{u\bar{d}}$ distribution and eventually play a role in the $W$ mass determination.

## 5.3 Exact cross sections versus effective approximations

When including all the tree-level diagrams for a four-fermion process in a computer program, one can loose some insight on which subsets of diagrams are really dominant and which are "sub-leading". On the other hand, in order to treat correctly the phase-space integrations and to get a reliable result, one should distinguish the main/secondary groups of diagrams. At the same time it is also useful to check the reliability of effective approximations that allow to evaluate given subsets of diagrams in a much simpler way. The natural way of forming subsets of diagrams is by isolating subgraphs that (with the in- and out- intermediate particles taken on mass shell) correspond to some gauge-invariant process of lowest order [41] (other ways of decomposition have not been successful, especially at high energies [42]). In this section, such a procedure is illustrated in the particular process $e^+e^- \to e^+e^-b\bar{b}$. This channel is important as a background for Higgs bosons searches. Figure 12 shows the 48 diagrams that make up the complete set (excluding the two that involve Higgs bosons): 8 multiperipheral, 16 invariance is discussed in the *WW Physics* Report. See also[40].

bremsstrahlung (single or non resonant, with a $\gamma/Z$ in the $t$-channel), 8 conversion (single- or double-resonant) and 16 annihilation (single- or non-resonant) graphs. The first three classes of diagrams involve the subprocesses $\gamma\gamma \to b\bar{b}$, $\gamma e \to Ve$ ($V = \gamma, Z$) and $e^+e^- \to VV$, respectively. The contribution of each subset to the total cross section has been computed exactly at tree level by CompHEP[43], and then compared with the corresponding results obtained through appropriate effective approximations that are described in the following.

Note that, in general, interferences between different subsets are found to be negligible at LEP2 energies, with the exception of the interferences of the bremsstrahlung diagrams with the $Z \to b\bar{b}$ decay, and the conversion diagrams with the $\gamma^* \to e^+e^-$ and $Z \to b\bar{b}$ decays (that gives -24 fb at $\sqrt{s} = 200$GeV). Then, apart from the interference between the bremsstrahlung diagram with $Z \to b\bar{b}$ and the one with $\gamma^* \to b\bar{b}$, which gives -3.2 fb at $\sqrt{s} = 200$GeV, all other interferences are found to be less than 1 fb at the same energy [41].

- **Effective approximation for multiperipheral diagrams.**

Using the equivalent photon spectrum in the Weizsäcker-Williams (WW) approximation [44], we can write the approximate formula for the total ‖ cross section corresponding to multiperipheral diagrams (first row in Fig. 12)

$$\sigma(\gamma\gamma \to b\bar{b}) = \int_{4m_b^2/s}^{1} dx_1 \int_{4m_b^2/x_1 s}^{1} dx_2\, \hat{\sigma}(\gamma\gamma \to b\bar{b})\, f_\gamma(x_1, \delta)\, f_\gamma(x_2, \delta) \tag{12}$$

where $f_\gamma(x, \delta)$ is given by [45]

$$f_\gamma(x, \delta) = \frac{\alpha}{2\pi}\left(\frac{1 + (1-x)^2}{x} \log\frac{1-x}{x^2}\frac{1}{\delta} - 2\frac{1-x}{x} + 2x\delta\right) \tag{13}$$

and $\delta = m_e^2/4m_b^2$. The subprocess cross section is given by (see, for instance, [46])

$$\hat{\sigma}(\gamma\gamma \to b\bar{b}) = \frac{2\alpha^2\pi}{27\hat{s}}\left((3 - v^4)\log\frac{1+v}{1-v} - 2v(2 - v^2)\right) \tag{14}$$

where $v = \sqrt{1 - 4m_b^2/\hat{s}}$. The results obtained through eq. (12) after a numerical integration are shown in Fig. 13 (dashed curve), and compared with the exact computation (solid curve) that includes also the multiperipheral $\gamma Z$- and $ZZ$-exchange diagrams (the last two are found to be suppressed by a factor $10^{-3}$ and $10^{-6}$, respectively, relative to the dominant $\gamma\gamma$ contribution). The agreement is excellent (indeed, the two curves overlap completely).

- **Effective approximation for $t$-channel photon exchange (bremsstrahlung) diagrams.**

Diagrams including the subprocess $\gamma^* e \to Ze$ (second row in Fig. 12) are well approximated by

$$\sigma(\gamma^* e \to Ze) = \int_{x_{min}}^{x_{max}} dx \int_{Q^2_{min}}^{Q^2_{max}} dQ^2 \frac{df_\gamma(x, Q^2)}{dQ^2} \hat{\sigma}(\gamma^* e \to Ze|Q^2) Br(Z \to b\bar{b}) \tag{15}$$

where

$$\frac{df_\gamma(x, Q^2)}{dQ^2} = \frac{\alpha}{2\pi}\left(\frac{1 + (1-x)^2}{xQ^2} - 2m_e^2 x \frac{1}{Q^4}\right) \tag{16}$$

‖ *i.e.* no cut on the invariant $b\bar{b}$ mass, $m_{b\bar{b}}$. The case including a cut on the invariant fermion mass and applications to the $m_{f\bar{f}}$ distribution are given below.

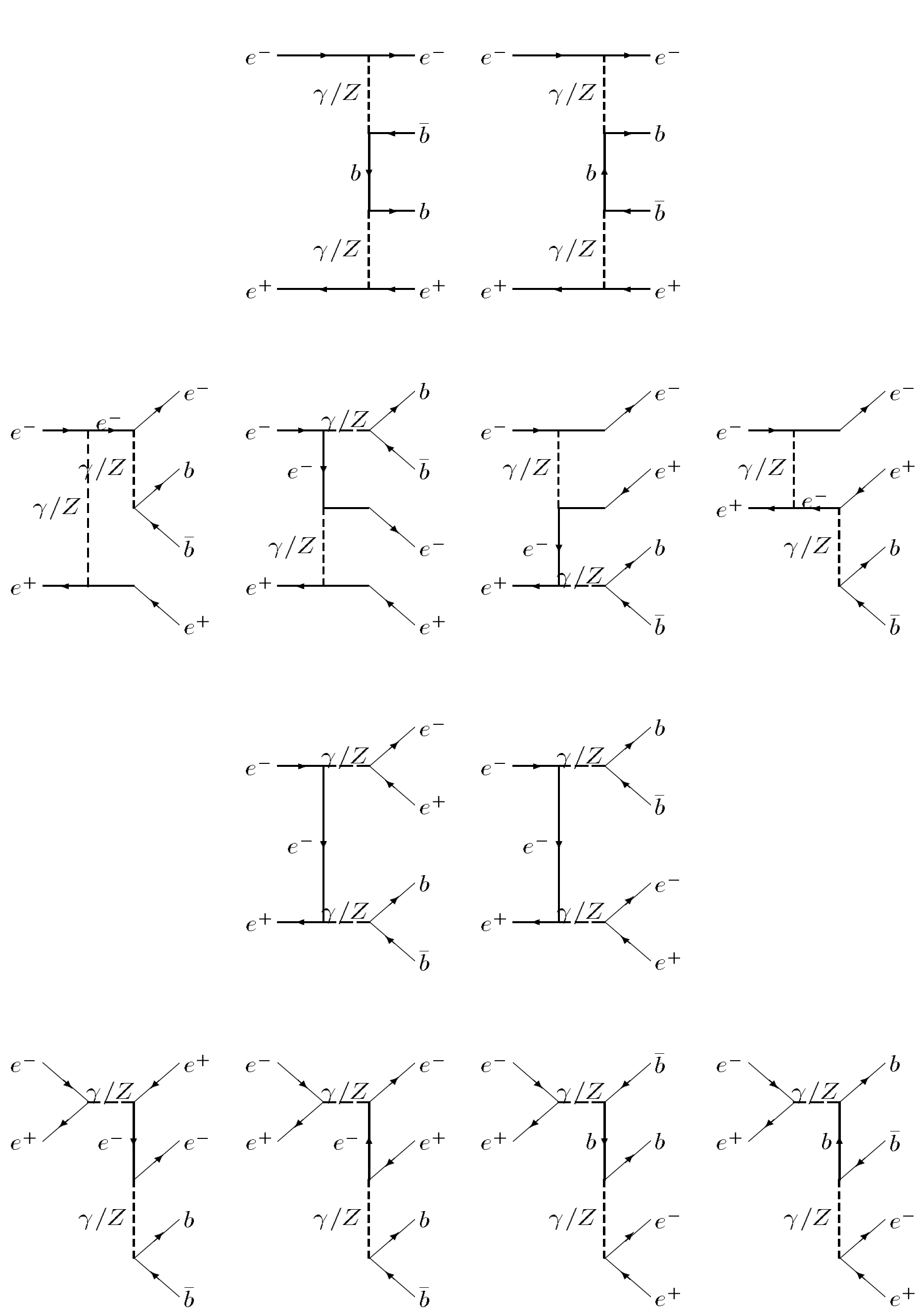

Figure 12: *Complete set of diagrams for the process* $e^+e^- \to e^+e^-b\bar{b}$.

with the integration limits

$$Q^2_{min} = m_e^2\frac{x^2}{1-x}, \qquad Q^2_{max} = m_Z^2 \tag{17}$$
$$x_{min} = \frac{(m_e+m_Z)^2}{s}, \qquad x_{max} = \frac{(\sqrt{s}-m_e)^2}{s}.$$

On the other hand, for the diagrams including the subprocess $\gamma^* e \to \gamma^* e$, one has

$$\sigma(\gamma^* e \to \gamma^* e) = \int_{x_{min}}^{x_{max}} dx \int_{Q^2_{min}}^{Q^2_{max}} dQ^2 \frac{1}{\pi} \int_{4m_b^2}^{(\sqrt{s}-m_e)^2} \frac{dM^2_{\gamma^*}}{M^3_{\gamma^*}} \frac{df_\gamma(x,Q^2)}{dQ^2} \hat{\sigma}(\gamma^* e \to \gamma^* e|Q^2)\Gamma(\gamma^* \to b\bar{b}) \tag{18}$$

with the integration limits

$$Q^2_{min} = m_e^2\frac{x^2}{1-x}, \qquad Q^2_{max} = 4m_b^2 \tag{19}$$
$$x_{min} = \frac{(m_e+2m_b)^2}{s}, \qquad x_{max} = \frac{(\sqrt{s}-m_e)^2}{s}.$$

The cross section for the subprocess $\gamma^* e \to Ve$, where $V$ denotes $Z$ or $\gamma^*$, can be written in the form

$$\begin{aligned}\hat{\sigma}(\gamma^* e \to Ve|Q^2) &= \frac{\alpha_e^2\pi}{\hat{s}} C_V \left(2(2x_V^2 - 2x_V + 1)\log\frac{\alpha+\beta}{\alpha-\beta} + \right. \\ &\left. +\beta\frac{x_e(7x_V+1) + x_\gamma x_V(3x_V^2 - 2x_V + 1)}{x_e + x_\gamma x_V(x_V - 1)} + \mathcal{O}(x_e, x_\gamma)\right)\end{aligned} \tag{20}$$

where

$$x_Z = m_Z^2/\hat{s}, \qquad x_{\gamma^*} = m^2_{\gamma^*}/\hat{s}$$
$$C_Z = \frac{8s_W^4 - 4s_W^2 + 1}{12 s_W^2 c_W^2}, \qquad C_{\gamma^*} = \tfrac{2}{3}$$

$$\alpha = 1 - x_V + x_e x_V - x_e^2 - x_\gamma(1 - x_e - x_V)$$
$$\beta = [(1 + (x_e - x_\gamma)^2 - 2x_e - 2x_\gamma)(1 + (x_e - x_V)^2 - 2x_e - 2x_V)]^{1/2}$$

$$x_e = m_e^2/\hat{s}, \quad x_\gamma = -Q_\gamma^2/\hat{s}, \quad \hat{s} = xs$$

• **Effective approximation for conversion, single and double-resonant diagrams.** We start from the conversion subprocess $e^+e^- \to \gamma^*\gamma^*$ (diagrams in the third row in Fig. 12). In this case

$$\sigma(e^+e^- \to \gamma_1^*(f_1\bar{f}_1)\gamma_2^*(f_2\bar{f}_2)) = \frac{1}{\pi^2}\int_{4m_{f_1}^2}^{(\sqrt{s}-2m_{f_2})^2} \frac{dM^2_{\gamma_1^*}}{M^3_{\gamma_1^*}} \int_{4m_{f_2}^2}^{(\sqrt{s}-M_{\gamma_1})^2} \frac{dM^2_{\gamma_2^*}}{M^3_{\gamma_2^*}}$$

$$\hat{\sigma}(e^+e^- \to \gamma^*\gamma^*)\Gamma(\gamma^* \to f_1\bar{f}_1)\Gamma(\gamma^* \to f_2\bar{f}_2) \tag{21}$$

where the off-shell photon decay width is given by

$$\Gamma(\gamma^* \to f\bar{f}) = \frac{\alpha}{3} Q_f^2 T_c M_{\gamma^*}(1+2x_f)\sqrt{1-4x_f} \tag{22}$$

$Q_f^2 = 1/9$ for the b-quark and 1 for the electron, $x_f = m_f^2/M_{\gamma^*}^2$. The color factor $T_c$ is equal to 3 for b-quark and 1 for the electron. The subprocess cross section is given by [47]

$$\hat{\sigma}(e^+e^- \to \gamma^*\gamma^*) = \frac{\alpha^2\pi}{s} C_D \left( A_D \log \frac{\alpha_D + \beta_D}{\alpha_D - \beta_D} - 3\alpha_D\beta_D \right) \tag{23}$$

where

$$\begin{aligned} C_D &= \frac{4}{1-x_1-x_2}, & A_D &= 1+(x_1+x_2)^2 \\ \alpha_D &= 1-x_1-x_2, & \beta_D &= \sqrt{1+(x_1-x_2)^2-2x_1-2x_2} \\ x_1 &= M_{\gamma_1^*}^2/s, & x_2 &= M_{\gamma_2^*}^2/s. \end{aligned}$$

For the single-resonant process $e^+e^- \to \gamma^* Z$ one has

$$\sigma(e^+e^- \to \gamma^*(f_1\bar{f}_1) + Z(f_2\bar{f}_2)) = \frac{1}{\pi}\int_{4m_{f_1}^2}^{\sqrt{s}} \frac{dM_{\gamma^*}^2}{M_{\gamma^*}^3} \hat{\sigma}(e^+e^- \to \gamma^*(f_1\bar{f}_1) + Z(f_2\bar{f}_2))$$
$$\Gamma(\gamma^* \to f_1\bar{f}_1) Br(Z \to f_2\bar{f}_2) \tag{24}$$

where the subprocess cross section is given by the formula eq. (23) with the parameters

$$\begin{aligned} C_D &= \frac{4}{1-x_{\gamma^*}-x_Z}\frac{8s_W^4-4s_W^2+1}{2s_W^2c_W^2}, & A_D &= 1+(x_{\gamma^*}+x_Z)^2 \\ \alpha_D &= 1-x_{\gamma^*}-x_Z, & \beta_D &= \sqrt{1+(x_{\gamma^*}-x_Z)^2-2x_{\gamma^*}-2x_Z} \\ x_{\gamma^*} &= M_{\gamma^*}^2/s, & x_Z &= m_Z^2/s. \end{aligned}$$

The cross section for the double resonant process $e^+e^- \to ZZ$, with the subsequent decays of $Z$ in the narrow width approximation, is given by the subprocess cross section eq. 23 with parameters

$$\begin{aligned} C_D &= \frac{38s_W^8-32s_W^6+24s_W^4-s_W^2+1}{16s_W^4c_W^4}\frac{1}{1-2x_Z}, & A_D &= 1+4x_Z^2 \\ \alpha_D &= 1-2x_Z, & \beta_D &= \sqrt{1-4x_Z}, \end{aligned}$$

multiplied by $Br(Z \to f_1\bar{f}_1)Br(Z \to f_2\bar{f}_2)$.

In figure 13, one can see that the exact computation (solid) is always reasonably recovered by the above approximations (dashes). Indeed, adding the approximate formulae for multi-peripheral, single and double conversion incoherently (with no interferences) the total cross section is reproduced within 5%.

It is also possible to improve on the approximation for the conversion diagrams that involves the $Z$, by including the finite-width effects and even the ISR, as we will discuss below.

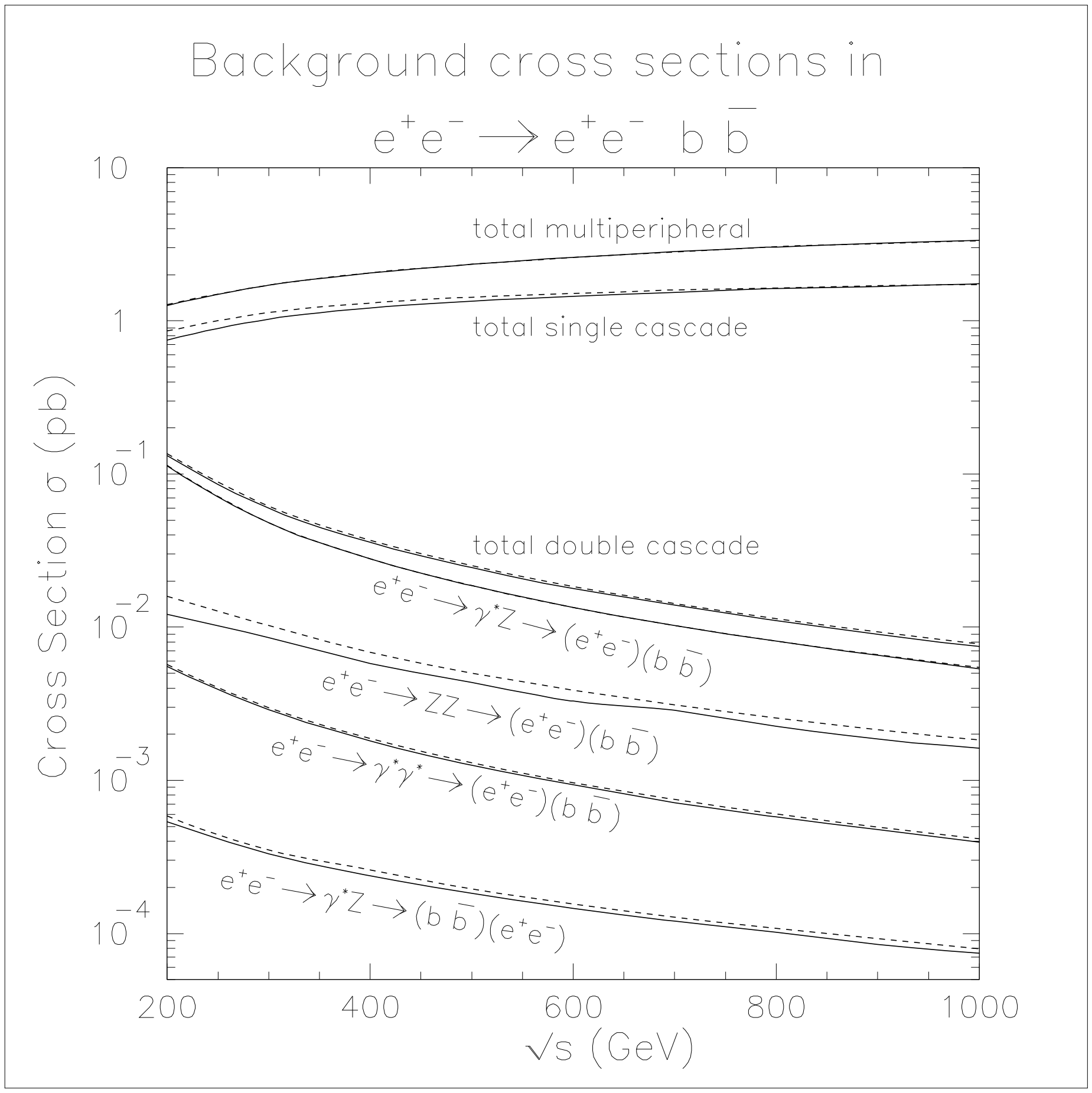

Figure 13: *Effective approximations (dashed lines) and exact calculations (solid lines) corresponding to various subsets of diagrams for the process* $e^+e^- \to e^+e^-b\bar{b}$.

## 5.4 Radiative corrections within the multiperipheral diagrams.

In this section, we discuss the accuracy of different versions of the Weizsäcker-Williams (WW) approximation [44] in describing both the integrated cross section (with a cut on the invariant mass of the fermions) as well as their $p_T$ distribution in two-photon processes. The effect of the QED corrections to the subprocess is also discussed within the approximation. In order to isolate the effect of the WW approximation error from other uncertainties (like QCD effects in two-photon hadron production), we study the $\gamma\gamma \to \tau^+\tau^-$ production as a reference process for more general cases.

Within the approximation the tree-level cross section is given by eq. (12), implemented with a cut on the invariant mass of the $\tau\tau$ pair. Several functions for the photon flux can be found in the literature, with the aim of giving more accurate descriptions of the exact rates. Indeed, it can happen that one formula can reproduce the total cross section quite precisely, but is less successful as far as some distributions are concerned, or *vice versa*. In general, the accuracy of a given approximation is both process and experimental-cut dependent.

Here, we compare how two different flux functions fare with the exact tree-level and one-loop

QED corrected result. This correction is only applied to the sub-process $\gamma\gamma \to \tau\tau$ [48]. The phase-space integration of the final state (7-dimensional for the 4-bodies and 10-dimensional for the 5-bodies) was performed by using the Monte Carlo integration package BASES [49].

The following two Weizsäcker-Williams spectra were examined

$$f_\gamma^{(1)}(x) = \frac{\alpha}{\pi x}\Big\{[1+(1-x)^2]\left(\ln\frac{2(1-x)E}{m}-\frac{1}{2}\right) -\frac{x^2}{2}(\ln x - 1) - \frac{(2-x)^2}{2}\ln(2-x)\Big\}, \tag{25}$$

$$f_\gamma^{(2)}(x) = \frac{\alpha}{\pi x}\left\{[1+(1-x)^2]\ln\left(\frac{2(1-x)}{x}\frac{E}{m}\right)-(1-x)\right\} \tag{26}$$

with $f_\gamma^{(1)}(x)$ [WWA(1)] and $f_\gamma^{(2)}(x)$ [WWA(2)] replacing $f_\gamma(x,\delta)$ in Eq.13 (note that the integration limits depends on the cut on $M_{\tau\tau}$ now).

| $\sigma$(pb) | Born | soft + loop | hard | $O(\alpha)$ corr. |
|---|---|---|---|---|
| exact | 6.017(6) | −2.361(2) | 2.403(2) | 0.70(5) |
| WWA(1) | 6.171(4) | −2.392(1) | 2.463(3) | 1.16(6) |
| WWA(2) | 8.370(6) | −3.224(2) | 3.316(4) | 1.10(5) |

Table 5: *Total cross section for $\tau$-pair production at $\sqrt{s} = 180GeV$ with the invariant-mass cut $M_{\tau\tau} > 30$ GeV. The photon contribution is separated into soft and hard at $k_\gamma = 1keV$. The last column shows the $\mathcal{O}(\alpha)$ correction in %.*

Table 5 summarizes the various components of the QED corrected total cross section calculated at $\sqrt{s} = 180$ GeV. The only kinematical cut applied is $M_{\tau\tau} > 30$GeV. The first spectrum, with $f_\gamma^{(1)}(x)$, reproduces the exact integrated cross section within 2% while the second choice overestimates the integrated cross section by almost 30%. Note that the $\mathcal{O}(\alpha)$ correction is small, about 1% and is reproduced in all three cases. The impact of the choice of the photon spectrum on the $p_T$ distribution for the the $\tau^-$ was also studied. The results are shown in Fig. 14. We observe that the first approximation reproduces nicely the exact distribution for small $p_T$ ($p_T < 20$GeV) while the second one is more suited in the medium $p_T$ range, though both fall down too fast in the large $p_T$ region (where, however, the statistics is very poor). From this example, one can conclude that the best choice of the non-leading term in the WW approximations depends on which quantity one wants to reproduce. For instance, the WWA(2) has been preferred in the analysis of $p_T$ distributions of two-photon process with high $p_T$ at TRISTAN [50].

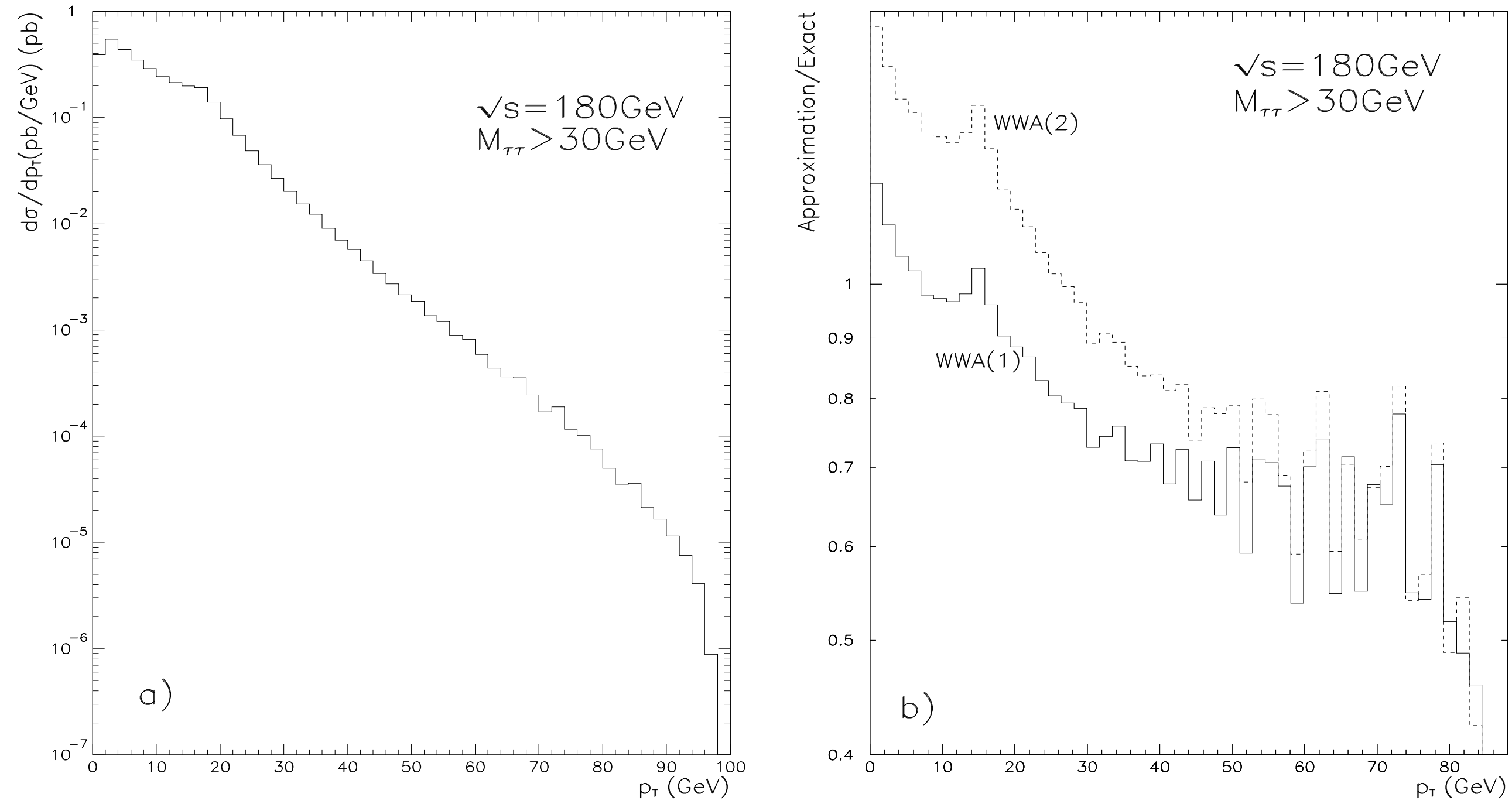

Figure 14: *a) $P_t$ distribution of $\tau^-$ at $\sqrt{s} = 180$ GeV for $M(\tau\tau) > 30$ GeV based on the exact calculation. b) shows the ratio of the WWA approximations over the exact result [see text for the definitions of WWA(1) and WWA(2)].*

## 5.5 Improved semi-analytical calculations for conversion-type four-fermion final states

We have already discussed how the conversion type diagrams can be approximated. The above approximations can be further improved by including finite-width effects and inserting ISR. In this sub-section, we report on four-fermion cross sections and invariant mass distributions as obtained by the *semi-analytical* method. All angular degrees of freedom in the phase space (five at tree level, seven if the ISR is included) are integrated analytically. After these analytical integrations, elegant and short expressions are obtained for invariant mass distributions. Fast, numerically stable, and highly precise numerical algorithms are then used to integrate the remaining phase-space degrees of freedom, namely the two or three squared invariant masses. Semi-analytical results are, however, not suitable for experimental simulations. In this sense, the semi-analytical and the Monte Carlo approach are complementary, and semi-analytical results may serve as benchmarks for numerical approaches, which usually rely on the Monte Carlo technique. A short review of semi-analytical calculations may be found in [51].

- **Convolution formulae at tree level**

In the framework of the semi-analytical technique, total four-fermion production tree-level cross

sections are given by

$$\sigma^{Born}(s) = \int ds_1 \int ds_2 \; \frac{\sqrt{\lambda}}{\pi s^2} \cdot \sum_k \frac{d^2\sigma_k(s, s_1, s_2)}{ds_1 ds_2} \quad . \tag{27}$$

Squared invariant masses for final-state fermion pair are represented by $s_1$ and $s_2$, and $\lambda \equiv \lambda(s, s_1, s_2)$ with $\lambda(a,b,c) = a^2+b^2+c^2-2ab-2ac-2bc$. The subscript index $k$ labels cross section contributions from squared amplitudes or interferences with distinct Feynman topologies *and* coupling structure. Partial double-differential cross sections have the form

$$\frac{d^2\sigma_k}{ds_1 ds_2} = \mathcal{C}_k(s, s_1, s_2) \cdot \mathcal{G}_k(s, s_1, s_2) \quad . \tag{28}$$

Coupling constants and off-shell boson propagators are collected in $\mathcal{C}_k$, while $\mathcal{G}_k$ is a kinematical function obtained after fivefold analytical integration over the angular phase-space variables. Both $\mathcal{C}_k$ and $\mathcal{G}_k$ are given by very compact expressions. For different charged current (CC) and neutral current (NC) processes, $\mathcal{C}_k$ and $\mathcal{G}_k$ may be found in references [51, 52, 53, 54, 55].

**•Complete $\mathcal{O}(\alpha)$ ISR with soft photon exponentiation**
A total four-fermion cross section with complete $\mathcal{O}(\alpha)$ ISR corrections including soft photon exponentiation is given by

$$\sigma^{ISR}(s) = \int ds_1 \int ds_2 \int\limits_{(\sqrt{s_1}+\sqrt{s_2})^2}^{s} \frac{ds'}{s} \sum_k \frac{d^3\Sigma_k(s, s'; s_1, s_2)}{ds_1 ds_2 ds'} \tag{29}$$

with the reduced squared center of mass energy $s'$ and

$$\frac{d^3\Sigma_k(s, s'; s_1, s_2)}{ds_1 ds_2 ds'} = \mathcal{C}_k(s', s_1, s_2) \cdot \left[\beta_e v^{\beta_e - 1} \mathcal{S}_k + \mathcal{H}_k\right] \quad , \tag{30}$$

where $\beta_e = 2\,\frac{\alpha}{\pi}\left[\ln(s/m_e^2) - 1\right]$ and $v = (1 - s'/s)$. Both the soft+virtual and hard contributions, $\mathcal{S}_k$ and $\mathcal{H}_k$, split into a universal, factorizing, process-independent and a non-universal, non-factorizing, process-dependent part. Using the twofold differential Born cross sections $\sigma_{k,0}(s'; s_1, s_2) \equiv \frac{\sqrt{\lambda}}{\pi s'^2} \cdot \mathcal{G}_k(s', s_1, s_2)$, one obtains

$$\begin{aligned} \mathcal{S}_k(s, s'; s_1, s_2) &= \left[1 + \bar{S}(s)\right] \sigma_{k,0}(s'; s_1, s_2) + \sigma_{\hat{S},k}(s'; s_1, s_2) \quad , \\ \mathcal{H}_k(s, s'; s_1, s_2) &= \underbrace{\bar{H}(s, s')\, \sigma_{k,0}(s'; s_1, s_2)}_{Universal\ Part} + \underbrace{\sigma_{\hat{H},k}(s, s'; s_1, s_2)}_{Non-universal\ Part} \end{aligned} \tag{31}$$

with the $\mathcal{O}(\alpha)$ soft+virtual and hard radiators $\bar{S}$ and $\bar{H}$ in the universal part given by

$$\bar{S}(s) = \frac{\alpha}{\pi}\left[\frac{\pi^2}{3} - \frac{1}{2}\right] + \frac{3}{4}\beta_e \qquad \bar{H}(s, s') = -\frac{1}{2}\left(1 + \frac{s'}{s}\right)\beta_e \quad . \tag{32}$$

If the index $k$ is associated with s-channel $e^+e^-$ annihilation diagrams only, non-universal ISR contributions are not present. Non-universal ISR contributions originate from the angular dependence of initial state $t$- and $u$-channel propagators. Since the non-universal cross section contributions $\sigma_{\hat{S},k}$ and $\sigma_{\hat{H},k}$ do not contain the large logarithm $\beta_e$, they only yield small cross section corrections up to a few percent. However, the analytical structure of $\sigma_{\hat{S},k}$ and $\sigma_{\hat{H},k}$ is very complex. An important feature of the non-universal corrections is the so-called *screening property*, i.e. an overall damping factor $s_1 \cdot s_2 / s^2$ in the non-universal corrections [52, 55, 56]. It is important to note that screening is a likely property with respect to the proper high energy unitarity behavior of the completely ISR corrected cross section. Semi-analytical treatments of complete ISR are presented in references [52, 55, 56]. Details of the non-universal contributions may be found in [56, 57].

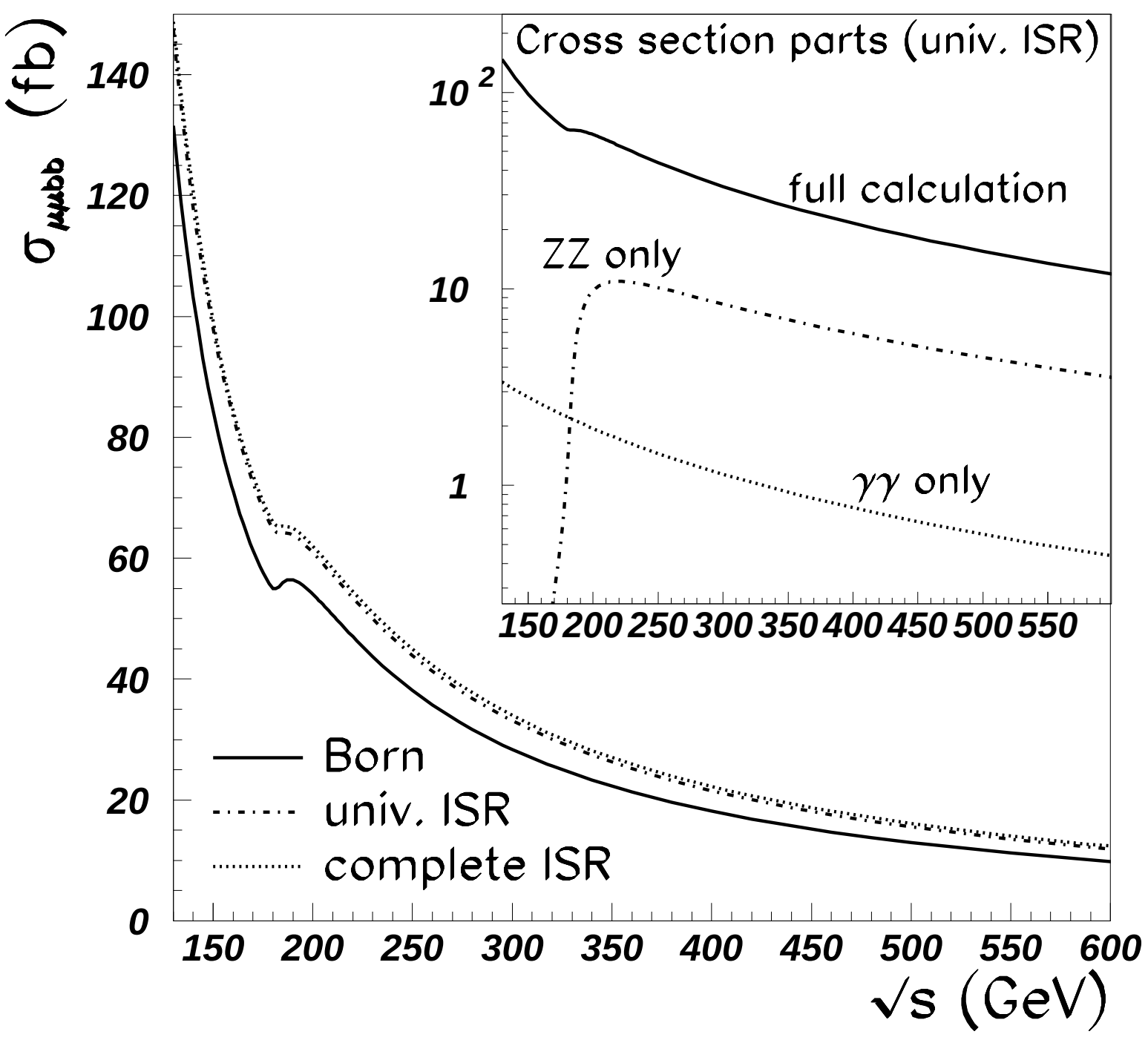

Figure 15: *The* `NC8` *cross section. The solid line represents the Born cross section, the dash-dotted line includes universal, and the dotted line includes all ISR corrections. In the inset, the universally ISR corrected* `NC8` *cross section is compared to the contributions from* $Z^0$ *and photon pair production.*

As an example for numerical results, figures 15 and 16 present total cross sections for the `NC8` process

$$e^+e^- \to (Z^0Z^0, Z^0\gamma, \gamma\gamma) \to \mu^+\mu^- \, b\bar{b} \tag{33}$$

without and with invariant fermion-pair mass cuts [56]. In figure 15, the cross section correction due to universal the ISR varies between 12% at $\sqrt{s}$=130 GeV and 21% at 600 GeV. The additional relative correction from the non-universal ISR increases from 9‰ at 130 GeV to 4.2% at 600 GeV. From figure 16 one can see how the `NC8` cross section approaches the cross section for the `NC2` reaction $e^+e^- \to (Z^0Z^0) \to \mu^+\mu^- \, b\bar{b}$ when invariant fermion-pair mass cuts are tightened. For the `NC2` reaction, the effect of universal ISR varies between –28% at the $Z^0$ pair threshold and approximately +10% at 600 GeV. Non-universal corrections to the NC2 reaction amount to less than half a percent below and around the threshold and rise to 1.5% at 600 GeV. Results for the NC24 process, that is with complete set of diagrams contributing to $e^+e^- \to f_1\bar{f}_1\, f_2\bar{f}_2$, with $f_1 \neq f_2 \neq e, \nu_e$, are found in reference [54] (see also below). Details of semi-analytical results for Higgs production and CC processes are reported by the working groups *Higgs*, *WW cross sections and distributions*, and *Event Generators for WW Physics* in this Report.

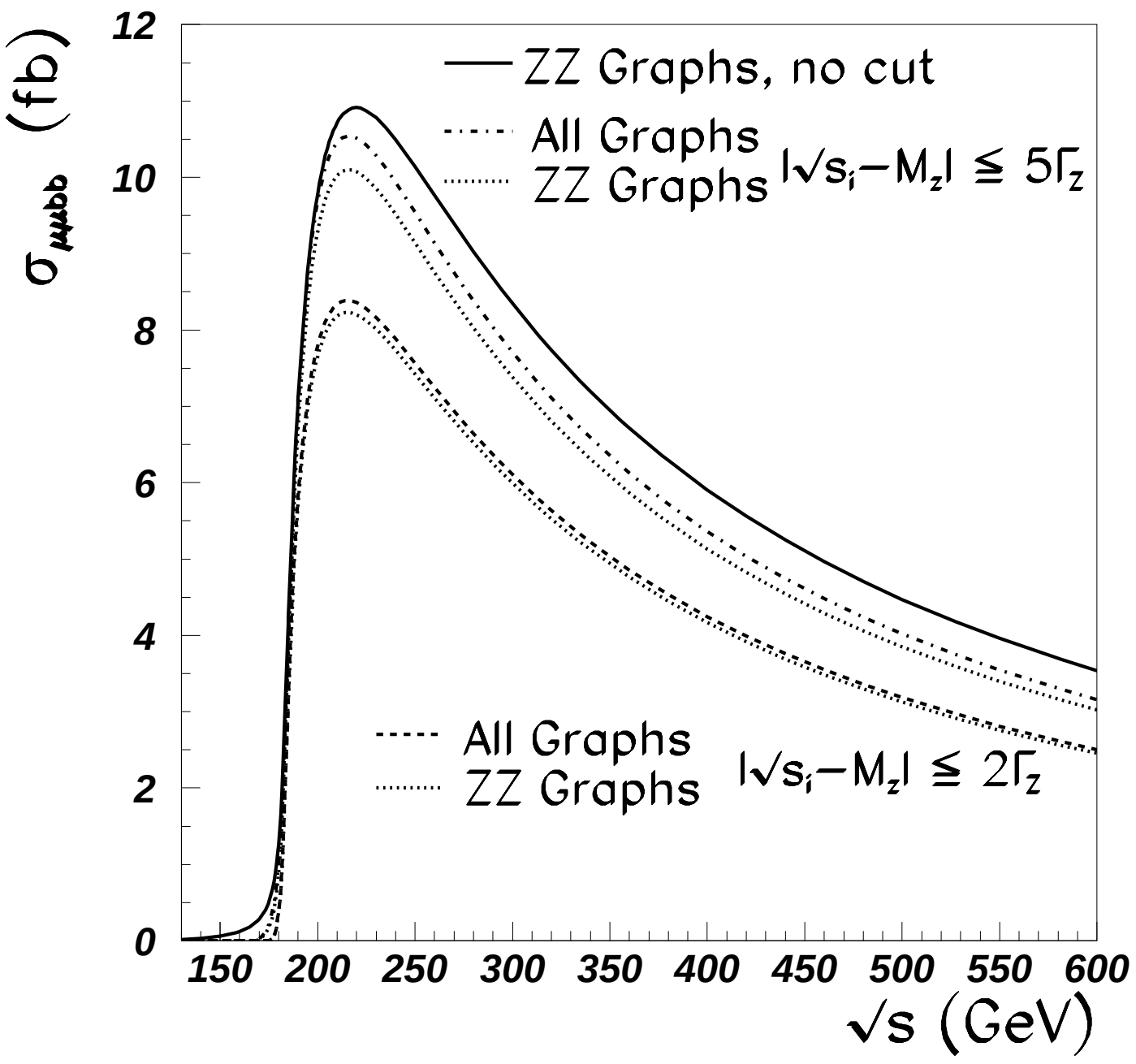

Figure 16: *The effect of cuts of* $2\cdot\Gamma_Z$ *and* $5\cdot\Gamma_Z$ *around the* $Z^0$ *mass* $M_Z$ *on the* **NC8** *('All Graphs') and* $Z^0$ *pair ('ZZ Graphs') cross sections. The cuts were applied to both the* $\mu^+\mu^-$ *and the* $b\bar{b}$ *pair invariant masses* $s_1$ *and* $s_2$*. All cross sections are universally ISR corrected.*

## 5.6 Cross sections for all four-fermion final states with inclusion of all diagrams

In this section, we report on the results of a study of the tree-level cross sections for all possible four-fermion final states, as listed in Tables 6-8. The complete set of diagrams is taken into account in each case (the corresponding total number of diagrams ($N_d$) is shown in the same tables). Higgs-boson contributions are not included. This comparative study involves seven codes: ALPHA [58], CompHEP [43], EXCALIBUR [59], grc4f (a package for computing four-fermion processes based on GRACE [38]), WWGENPV/HIGGSPV [60], WPHACT [61] and WTO [62]. For a detailed description of the codes see the *Event Generators for WW Physics* Report. In this comparison ISR and gluon-exchange diagrams for the hadronic four-

| NN | $e^+e^- \to$ | $N_d$ | ALPHA | CompHEP | EXCALIBUR | grc4f | HIGGSPV* | WPHACT | WTO |
|---|---|---|---|---|---|---|---|---|---|
| 1 | $e^+e^-\nu_e\bar{\nu}_e$ | 56 | 257.3(2) | 255.4(13) | 256.7(2) | 256.8(7) | — | 257.0(2) | — |
| 2<br>3<br>4<br>5 | $e^-\bar{\nu}_e\nu_\mu\mu^+$<br>$e^-\bar{\nu}_e\nu_\tau\tau^+$<br>$\nu_e e^+\mu^-\bar{\nu}_\mu$<br>$\nu_e e^+\tau^-\bar{\nu}_\tau$ | 18 | 227.1(1) | 227.8(5) | 227.2(1) | 227.0(2) | 226.9(4)* | 227.3(1) | 227.2(1) |
| 6<br><br>7 | $\mu^+\mu^-\nu_\mu\bar{\nu}_\mu$<br><br>$\tau^+\tau^-\nu_\tau\bar{\nu}_\tau$ | 19 | 228.6(2) | 227.3(8) | 228.6(2) | 228.7(7)<br>228.3(2)[m]<br>225.1(4)[m] | — | 228.6(0) | 228.6(1) |
| 8<br><br>9 | $\mu^-\bar{\nu}_\mu\nu_\tau\tau^+$<br><br>$\tau^-\bar{\nu}_\tau\nu_\mu\mu^+$ | 9 | 218.5(1) | 218.4(4) | 218.2(1) | 218.5(2)<br>218.3(2)[m] | 218.4(1)* | 218.6(2) | 218.1(0) |
| 10 | $e^+e^-e^+e^-$ | 144 | — | — | 109.7(2) | 109.0(6) | — | 109.6(2) | — |
| 11<br><br>12 | $e^+e^-\mu^+\mu^-$<br><br>$e^+e^-\tau^+\tau^-$ | 48 | — | 113.1(15) | 116.6(2) | 116.5(3)<br>111.6(1)[m]<br>58.68(5)[m] | 112.8(19) | 116.8(2) | — |
| 13<br><br>14 | $\mu^+\mu^-\mu^+\mu^-$<br><br>$\tau^+\tau^-\tau^+\tau^-$ | 48 | 5.456(5) | 5.439(32) | 5.476(10) | 5.467(9)<br>5.387(7)[m]<br>3.786(3)[m] | 5.65(52) | 5.472(5) | 5.460(17) |
| 15 | $\mu^+\mu^-\tau^+\tau^-$ | 24 | 11.00(1)<br>9.25(1)[m] | 10.95(4) | 10.99(2) | 10.97(4)<br>9.233(16)[m] | 11.01(1) | 11.02(2) | 11.00(1) |
| 16<br>17 | $e^+e^-\nu_\mu\bar{\nu}_\mu$<br>$e^+e^-\nu_\tau\bar{\nu}_\tau$ | 20 | — | 14.13(4) | 14.15(2) | 14.14(3) | 14.34(17) | 14.16(1) | — |
| 18<br><br>19 | $\nu_e\bar{\nu}_e\mu^+\mu^-$<br><br>$\nu_e\bar{\nu}_e\tau^+\tau^-$ | 19 | 17.78(2) | 17.78(5) | 17.92(4) | 17.75(3)<br>17.39(3)[m]<br>11.08(1)[m] | 17.79(1) | 17.81(1) | 17.83(15) |
| 20<br><br>21 | $\nu_\tau\bar{\nu}_\tau\mu^+\mu^-$<br><br>$\nu_\mu\bar{\nu}_\mu\tau^+\tau^-$ | 10 | 10.10(1) | 10.09(3) | 10.14(2) | 10.10(3)<br>10.038(8)[m]<br>8.533(6)[m] | 10.10(1) | 10.09(2) | 10.05(3) |
| 22 | $\nu_e\bar{\nu}_e\nu_e\bar{\nu}_e$ | 36 | 4.091(2) | 4.108(22) | 4.087(2) | 4.085(5) | — | 4.089(1) | — |
| 23<br>24 | $\nu_e\bar{\nu}_e\nu_\mu\bar{\nu}_\mu$<br>$\nu_e\bar{\nu}_e\nu_\tau\bar{\nu}_\tau$ | 12 | 8.335(4) | 8.335(9) | 8.335(3) | 8.335(6) | 8.369(54) | 8.339(1) | 8.356(2) |
| 25<br>26 | $\nu_\mu\bar{\nu}_\mu\nu_\mu\bar{\nu}_\mu$<br>$\nu_\tau\bar{\nu}_\tau\nu_\tau\bar{\nu}_\tau$ | 12 | 4.065(4) | 4.107(8) | 4.071(1) | 4.063(4) | 4.067(7) | 4.068(1) | 4.117(1) |
| 27 | $\nu_\mu\bar{\nu}_\mu\nu_\tau\bar{\nu}_\tau$ | 6 | 8.245(4) | 8.234(9) | 8.240(3) | 8.240(4) | 8.237(6) | 8.241(1) | 8.241(1) |

Table 6: *Cross sections (in fb) for all the leptonic four-fermion final states. The superscript* $[m]$ *marks all the results where complete fermion-mass effects are taken into account. The asterisks in the HIGGSPV column distinguish cross sections computed by the WWGENPV version of the program.*

| NN | $e^+e^- \to$ | $N_d$ | ALPHA | CompHEP | EXCALIBUR | grc4f | HIGGSPV* | WPHACT | WTO |
|---|---|---|---|---|---|---|---|---|---|
| 1 | $e^-\bar{\nu}_e u\bar{d}$ | 20 | 692.9(5) | 693.3(13) | 692.8(4) | 692.5(4) | 691.9(12)* | 692.7(5) | 692.8(3) |
| 2 | $e^-\bar{\nu}_e c\bar{s}$ | | | | | 692.1(5)[m] | | | |
| 3 | $\nu_e e^+ d\bar{u}$ | | | | | | | | |
| 4 | $\nu_e e^+ s\bar{c}$ | | | | | | | | |
| 5 | $\mu^-\bar{\nu}_\mu u\bar{d}$ | 10 | 666.3(4) | 664.9(11) | 666.9(4) | 666.2(4) | 666.8(5)* | 666.7(4) | 666.2(1) |
| 6 | $\mu^-\bar{\nu}_\mu c\bar{s}$ | | | | | 665.7(4)[m] | | | |
| 7 | $\mu^+\nu_\mu d\bar{u}$ | | | | | | | | |
| 8 | $\mu^+\nu_\mu s\bar{c}$ | | | | | | | | |
| 9 | $\tau^-\bar{\nu}_\tau u\bar{d}$ | | | | | 665.7(4)[m] | | | |
| 10 | $\tau^-\bar{\nu}_\tau c\bar{s}$ | | | | | 665.3(4)[m] | | | |
| 11 | $\tau^+\nu_\tau d\bar{u}$ | | | | | | | | |
| 12 | $\tau^+\nu_\tau s\bar{c}$ | | | | | | | | |
| 13 | $e^+e^- u\bar{u}$ | 48 | — | 85.78(63) | 86.87(9) | 86.88(9) | 84.91(93) | 86.80(15) | 87.64(34) |
| 14 | $e^+e^- c\bar{c}$ | | | | | 78.20(42)[m] | | | |
| 15 | $e^+e^- d\bar{d}$ | 48 | — | 42.77(21) | 43.05(5) | 42.95(7) | 43.61(41) | 43.01(9) | 43.35(23) |
| 16 | $e^+e^- s\bar{s}$ | | | | | | | | |
| 17 | $e^+e^- b\bar{b}$ | | | | | 36.51(5)[m] | | | |
| 18 | $\mu^+\mu^- u\bar{u}$ | 24 | 24.71(2) | 24.58(6) | 24.80(3) | 24.69(3) | 24.68(1) | 24.69(2) | 24.59(4) |
| | | | 24.48(3)[m] | | | 24.53(3)[m] | | | |
| 19 | $\mu^+\mu^- c\bar{c}$ | | | | | 24.57(6)[m] | | | |
| 20 | $\tau^+\tau^- u\bar{u}$ | | | | | 20.29(3)[m] | | | |
| 21 | $\tau^+\tau^- c\bar{c}$ | | | | | 20.39(5)[m] | | | |
| 22 | $\mu^+\mu^- d\bar{d}$ | 24 | 23.74(2) | 23.65(7) | 23.70(4) | 23.71(1) | 23.73(1) | 23.71(2) | 23.58(5) |
| | | | | | | 23.60(1)[m] | | | |
| 23 | $\mu^+\mu^- s\bar{s}$ | | | | | | | | |
| 24 | $\mu^+\mu^- b\bar{b}$ | | | | | 22.98(3)[m] | | | |
| 25 | $\tau^+\tau^- d\bar{d}$ | | | | | 20.03(3)[m] | | | |
| 26 | $\tau^+\tau^- s\bar{s}$ | | | | | | | | |
| 27 | $\tau^+\tau^- b\bar{b}$ | | | | | 19.49(2)[m] | | | |
| 28 | $\nu_e\bar{\nu}_e u\bar{u}$ | 19 | 23.89(2) | 23.88(5) | 23.89(1) | 23.82(4) | 23.95(5) | 23.87(1) | 24.02(14) |
| 29 | $\nu_e\bar{\nu}_e c\bar{c}$ | | | | | 24.26(3)[m] | | | |
| 30 | $\nu_e\bar{\nu}_e d\bar{d}$ | 19 | 20.66(2) | 20.62(5) | 20.67(1) | 20.63(2) | 20.67(8) | 20.65(1) | 20.68(4) |
| 31 | $\nu_e\bar{\nu}_e s\bar{s}$ | | | | | | | | |
| 32 | $\nu_e\bar{\nu}_e b\bar{b}$ | | | | | 19.63(2)[m] | | | |
| 33 | $\nu_\mu\bar{\nu}_\mu u\bar{u}$ | 10 | 21.04(2) | 21.07(3) | 21.09(1) | 21.07(2) | 21.08(1) | 21.09(1) | 21.13(14) |
| 34 | $\nu_\mu\bar{\nu}_\mu c\bar{c}$ | | | | | 21.32(2)[m] | | | |
| 35 | $\nu_\tau\bar{\nu}_\tau u\bar{u}$ | | | | | | | | |
| 36 | $\nu_\tau\bar{\nu}_\tau c\bar{c}$ | | | | | | | | |
| 37 | $\nu_\mu\bar{\nu}_\mu d\bar{d}$ | 10 | 19.88(2) | 19.80 (4) | 19.86(1) | 19.85(2) | 19.86(1) | 19.87(1) | 19.89(4) |
| 38 | $\nu_\mu\bar{\nu}_\mu s\bar{s}$ | | | | | | | | |
| 39 | $\nu_\mu\bar{\nu}_\mu b\bar{b}$ | | | | | 19.16(1)[m] | | | |
| 40 | $\nu_\tau\bar{\nu}_\tau d\bar{d}$ | | | | | | | | |
| 41 | $\nu_\tau\bar{\nu}_\tau s\bar{s}$ | | | | | | | | |
| 42 | $\nu_\tau\bar{\nu}_\tau b\bar{b}$ | | | | | | | | |

Table 7: *Cross sections (in fb) for all the semileptonic four-fermion final states. The notation is the same as in Table 6.*

fermion final states (when implemented) are switched off. The effect of non-zero fermion masses for some of the processes has also been investigated by ALPHA and grc4f (see Tables 6-8). Total cross sections have been computed at the centre-of-mass energy $\sqrt{s} = 190\text{GeV}$, with the following cuts: $E_{\ell^\pm} > 1\text{GeV}$ , $E_q > 3\text{GeV}$ , $\theta(\ell^\pm - beam) > 10^o$ , $\theta(\ell^\pm - \ell'^\pm) > 5^o$ , $\theta(\ell^\pm - q) > 5^o$ , $M_{qq^{(\prime)}} > 5\text{GeV}$ (cuts on the fermion energy variables are loosened in the case of massive fermions). Furthermore, in order to better check the agreement among the different codes, a *canonical* set of input parameter has been agreed upon in all the computations, that is $M_Z = 91.1888\text{GeV}$, $\Gamma_Z = 2.4974\text{GeV}$ , $M_W = 80.23\text{GeV}$ , $\Gamma_W = \frac{3G_F M_W^3}{\sqrt{8}\pi} = 2.0337\text{GeV}$ , $\alpha^{-1}(2M_W) = 128.07$, $G_F = 1.16639\ 10^{-5}\text{GeV}^{-2}$ , $\sin^2\theta_W$ from $\frac{\alpha(2M_W)}{2\sin^2\theta_W} = \frac{G_F M_W^2}{\pi\sqrt{2}}$. In Table 6, the cross sections for all the four-lepton final states are shown, in Table 7 the ones for the semileptonic states and in Table 8 the ones for the hadronic four-fermion states. The error in the last one or two digits, corresponding to the Monte Carlo event generator, is also shown in parenthesis. One can see that the agreement among the different central values is in general at the level of a few per-mil, and even better in some cases. Note that, with the cuts above, the effect of the fermion masses can be not negligible, as can be seen by comparing the rates for muons to those for $\tau$'s for instance, (*cf.* Tables 6-7).

| NN | $e^+e^- \to$ | $N_d$ | ALPHA | CompHEP | EXCALIBUR | grc4f | HIGGSPV* | WPHACT | WTO |
|---|---|---|---|---|---|---|---|---|---|
| 1<br>2 | $u\bar{u}d\bar{d}$<br>$c\bar{c}s\bar{s}$ | 35 | 2063(1) | 2045(7) | 2064(1) | 2064(3)<br>2063(3)[m] | — | 2064(0) | 2062(1) |
| 3<br><br>4 | $u\bar{d}s\bar{c}$<br><br>$d\bar{u}c\bar{s}$ | 11 | 2015(1) | 2019(6) | 2015(1) | 2015(1)<br>2013(3)[m] | 2015(1)* | 2015(1) | 2014(0) |
| 5<br>6 | $u\bar{u}u\bar{u}$<br>$c\bar{c}c\bar{c}$ | 48 | 25.65(3) | — | 25.75(1) | 25.58(8)<br>26.36(3)[m] | 25.36(17) | 25.73(1) | — |
| 7<br>8<br>9 | $d\bar{d}d\bar{d}$<br>$s\bar{s}s\bar{s}$<br>$b\bar{b}b\bar{b}$ | 48 | 23.49(2) | — | 23.49(1) | 23.49(8)<br><br>22.11(11)[m] | 23.28(14) | 23.49(1) | — |
| 10 | $u\bar{u}c\bar{c}$ | 24 | 51.54(5)<br>52.21(5)[m] | 51.58(10) | 51.59(2) | 51.57(3)<br>52.28(3)[m] | 51.60(4) | 51.64(5) | 51.50(7) |
| 11<br>12<br>13<br>14 | $u\bar{u}s\bar{s}$<br>$u\bar{u}b\bar{b}$<br>$c\bar{c}d\bar{d}$<br>$c\bar{c}b\bar{b}$ | 24 | 49.58(5) | 49.47(14) | 49.69(1) | 49.68(4)<br>48.68(5)[m]<br>50.35(7)[m]<br>49.29(5)[m] | 49.71(4) | 49.66(3) | 49.67(11) |
| 15<br>16<br>17 | $d\bar{d}s\bar{s}$<br>$d\bar{d}b\bar{b}$<br>$s\bar{s}b\bar{b}$ | 24 | 47.04(5) | 46.95(9) | 47.12(2) | 47.12(3)<br>46.08(4)[m] | 47.11(6) | 47.11(3) | 47.11(9) |

Table 8: *Cross sections (in fb) for all the hadronic four-fermion final states. The notation is the same as in Table 6.*

# 6 Three Vector-Boson Production

LEP2 can in principle be sensitive to quartic self-interactions of the electroweak vector bosons, through the production of two bosons plus one large-angle hard photon in the channels $e^+e^- \to WW\gamma$, $ZZ\gamma$ and $Z\gamma\gamma$. While the inclusion of quartic couplings is essential to maintain gauge invariance, these couplings cannot be simply isolated as subtle cancellations among many diagrams, including also trilinear couplings, take place. Nevertheless, triple vector-boson production can be used as a test for the presence of anomalous couplings, in particular $\gamma\gamma WW$ $\gamma ZWW$ and $\gamma\gamma ZZ$ [63].

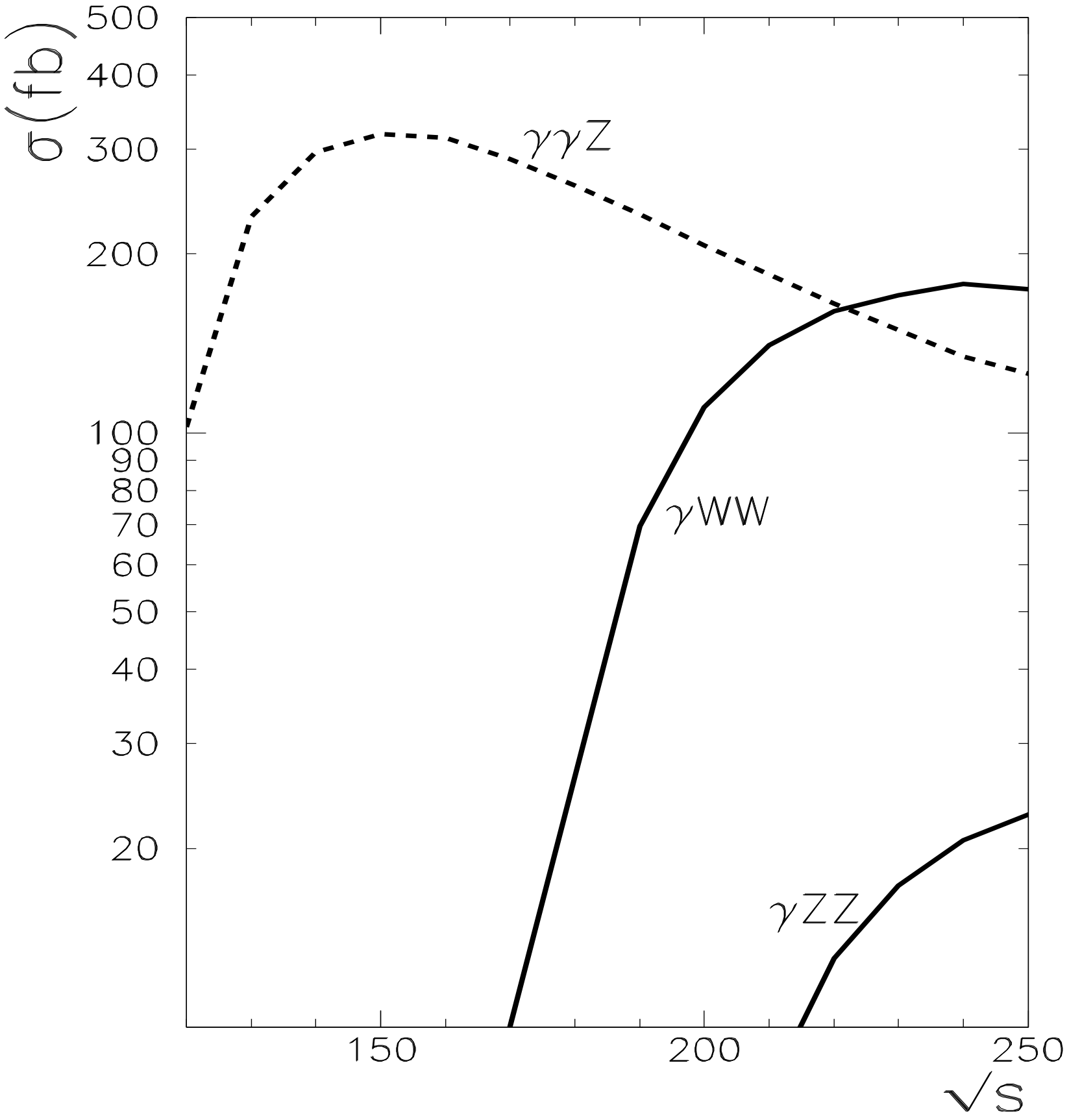

Figure 17: *Three vector-bosons cross sections. The applied cuts are* $\cos\theta_{eV} > 15^o$ *(*$V = W^{\pm}, Z\gamma$*) and* $\cos\theta_{VV} > 10^o$*, as well as a cut on* $p_T^{\gamma} > 10$ *GeV.*

The cross-sections for the production of three vector bosons are shown in Figure 17, where (generous) angular cuts $\cos\theta_{eV} > 15^o$ ($V = W^{\pm}, Z\gamma$) and $\cos\theta_{VV} > 10^o$, as well as a cut on $p_T^{\gamma} > 10$ GeV, have been imposed to avoid backgrounds. The $WW\gamma$ cross section increases very sharply near 170GeV (just above threshold) but LEP2 has barely enough energy to produce these final states with healthy statistics. One must therefore strive for the highest possible energy in order to increase the statistics. Furthermore, the sensitivity to anomalous couplings also rises with energy. Estimates using the above cuts have shown that even with a centre-of-mass energy of 230 GeV, one would need a two-orders-of-magnitude increase in precision to reach the level needed to test New Physics.

**Acknowledgments:**
We would like to thank Alain Blondel and Tim Stelzer for briefing us at the early stages of the Workshop and taking an active part in our meetings and Wim Beenakker for checking the numbers in Table 1. Misha Dubinin is grateful to the Minami-Tateya group (Computing Physics Division, KEK, Tsukuba) for giving him the possibility to use their computing facilities. The conveners gratefully acknowledge the unfailing cooperation of Dima Bardin as concerns 4-fermion processes.
This work has been partly supported by the *Human Capital and Mobility Program* of the European Community (contract numbers: ERB-CHRX-CT92-0004, ERB-CHRX-CT93-0132, ERB-CHRX-CT93-0319 and ERB-CHRX-CT93-0357); *INTAS* (93-1180 and 93-744); the *International Science Foundation* (M9B000 and M9B300) and the US *Departement of Energy* (DE–FG0295ER40896).